\documentclass[journal]{IEEEtran}

\usepackage{color}
\usepackage{lineno}
\usepackage{cite}
\usepackage{verbatim}
\usepackage{amsmath,amssymb,amsfonts}
\usepackage{algorithmic}
\usepackage{graphicx}
\usepackage{textcomp}
\usepackage{xcolor}
\usepackage{subfigure}
\usepackage{setspace}
\usepackage{svg}
\usepackage[ruled]{algorithm2e}
\usepackage{booktabs}
\usepackage{url}
\usepackage{diagbox}
\usepackage{pifont}
\usepackage{xcolor}
\usepackage{multicol}
\usepackage{multirow}
\usepackage{tikz}
\usepackage[colorlinks=true,linkcolor=blue,citecolor=blue,urlcolor=blue]{hyperref}
\definecolor{lime}{HTML}{A6CE39}
\DeclareRobustCommand{\orcidicon}{%
    \begin{tikzpicture}
    \draw[lime, fill=lime] (0,0) 
    circle [radius=0.16] 
    node[white] {{\fontfamily{qag}\selectfont \tiny ID}};    \draw[white, fill=white] (-0.0625,0.095) 
    circle [radius=0.007];    \end{tikzpicture}
    \hspace{-2mm}}
\foreach \x in {A, ..., Z}{%
    \expandafter\xdef\csname orcid\x\endcsname{\noexpand\href{https://orcid.org/\csname orcidauthor\x\endcsname}{\noexpand\orcidicon}}
    }
\usepackage{nomencl}
\SetCommentSty{mycommfont}

\SetKwInput{KwInput}{Input}                
\SetKwInput{KwOutput}{Output}              

\begin{document}



\title{Through-the-Wall Radar Human Activity Micro-Doppler Signature Representation Method Based on Joint Boulic-Sinusoidal Pendulum Model\\
\thanks{Manuscript received January 15th, 2024; revised XXXXXXXX XXth, 2024; accepted XXXXXXXX XXth, 2024. Date of publication XXXXXXXX XXth, 2024; date of current version XXXXXXXX XXth, 2024. This work was supported in part by the National Natural Science Foundation of China under Grant 62101042. This work was also supported in part by Beijing Institute of Technology Research Fund Program for Young Scholars under Grant XSQD-202205005. (Corresponding author: Xiaodong Qu.)\par
Xiaopeng Yang, Zeyu Ma, and Hao Zhang, are with the School of Information and Electronics, Beijing Institute of Technology, Beijing 100081, China, and also with the Jiaxing Research Center of Beijing Institute of Technology, Jiaxing 314000, China (email: xiaopengyang@bit.edu.cn; 3220220653@bit.edu.cn; 3220220662@bit.edu.cn).\par
Weicheng Gao, and Xiaodong Qu, are with the School of Information and Electronics, Beijing Institute of Technology, Beijing 100081, China, and with the Key Laboratory of Electronic and Information Technology in Satellite Navigation, Beijing Institute of Technology, Beijing 100081, China (e-mail: JoeyBG@126.com; xdqu@bit.edu.cn).\par
Digital Object Identifier 10.1109/TMTT.2024.XXXXXXX\par}}

\author{Xiaopeng~Yang\orcidA{},~\IEEEmembership{Senior~Member,~IEEE,}   
        Weicheng~Gao\orcidB{},~\IEEEmembership{Graduate~Student~Member,~IEEE,}
        Xiaodong~Qu\orcidC{},~\IEEEmembership{Member,~IEEE,}
        Zeyu~Ma\orcidD{},~\IEEEmembership{Student~Member,~IEEE}
        and~Hao~Zhang\orcidE{},~\IEEEmembership{Student~Member,~IEEE}      
        \vspace{-0.2cm}
        }
        
\markboth{IEEE TRANSACTIONS ON MICROWAVE THEORY AND TECHNIQUES, VOL. XX, 2024}%
{Shell \MakeLowercase{\textit{et al.}}: Bare Demo of IEEEtran.cls for IEEE Journals}

\maketitle

\begin{abstract}
With the help of micro-Doppler signature, ultra-wideband (UWB) through-the-wall radar (TWR) enables the reconstruction of range and velocity information of limb nodes to accurately identify indoor human activities. However, existing methods are usually trained and validated directly using range-time maps (RTM) and Doppler-time maps (DTM), which have high feature redundancy and poor generalization ability. In order to solve this problem, this paper proposes a human activity micro-Doppler signature representation method based on joint Boulic-sinusoidal pendulum motion model. In detail, this paper presents a simplified joint Boulic-sinusoidal pendulum human motion model by taking head, torso, both hands and feet into consideration improved from Boulic-Thalmann kinematic model. The paper also calculates the minimum number of key points needed to describe the Doppler and micro-Doppler information sufficiently. Both numerical simulations and experiments are conducted to verify the effectiveness. The results demonstrate that the proposed number of key points of micro-Doppler signature can precisely represent the indoor human limb node motion characteristics, and substantially improve the generalization capability of the existing methods for different testers.\par
\end{abstract}

\begin{IEEEkeywords}
through-the-wall radar, human activity recognition, micro-Doppler signature, feature extraction, corner representation.
\end{IEEEkeywords}

\IEEEpeerreviewmaketitle

\makenomenclature
\nomenclature{AEN}{Autoencoders Network}
\nomenclature{ResNet}{Residual Convolutional Neural Network}
\nomenclature{AI}{Artificial Intelligence}
\nomenclature{mmWave}{Millimeter-Wave}
\nomenclature{T-F}{Time-Frequency}
\nomenclature{HAR}{Human Activity Recognition}
\nomenclature{LSTM}{Long-Short Term Memory}
\nomenclature{UWB}{Ultra-Wideband}
\nomenclature{RGB}{Red Green Blue}
\nomenclature{CPI}{Coherent Integration Interval}
\nomenclature{PRI}{Pulse Repetition Intervals}
\nomenclature{LFMCW}{Linear Frequency Modulated Continuous Wave}
\nomenclature{RTM}{Rage-Time Map}
\nomenclature{DTM}{Doppler-Time Map}
\nomenclature{MTI}{Moving Target Indication}
\nomenclature{STFT}{Short-Time Fourier Transform}
\nomenclature{FOGDD}{First-Order Generalized Gaussian Directional Derivative}
\nomenclature{SOGGDD}{Second-Order Generalized Gaussian Directional Derivative}
\nomenclature{SODDC}{Second-Order Directional Derivative Correlation}
\nomenclature{MNCP}{Minimum Number of Corner Points}
\nomenclature{UCL}{University College London}
\nomenclature{TUDelft}{Delft University of Technology}
\nomenclature{SNR}{Signal-to-Noise Ratio}
\nomenclature{PSNR}{Peak Signal-to-Noise Ratio}
\nomenclature{T-SNE}{T-Distributed Stochastic Neighbor Embedding}
\nomenclature{TWR}{Through-the-Wall Radar}
\nomenclature{NMS}{Non-Maximum Suppression}
\nomenclature{1/2/3D}{One/Two/Three-Dimensional}
\nomenclature{EMD}{Empirical Modal Decomposition}
\nomenclature{E-M-Dist}{Earth Mover's Distance (Wassertein Distance)}
\nomenclature{OMP}{Orthogonal Matching Pursuit}
\printnomenclature

\section{Introduction}
\IEEEPARstart{I}{ndoor} human activity recognition has gradually become a popular research topic in the field of radar and intelligent signal interpretation \cite{Amin1,TMTT1,TMTT2,TSP,Amin2,VCChen}. In the literature, contact devices, cameras, mmWave radar, and WIFI were devoted for activity recognition \cite{TMTT1,Radar VS Camera}. Recently, UWB TWR was also employed, which could penetrate the wall and work under complex environments, avoiding privacy and security issues, and has been used in different applications \cite{Guolong Cui, Tian Jin, Songyongping, Jianqi Wang, Shengbo Ye, Yang Yang, Yipeng Ding, Jiahui Chen, Yutao Xiang}.\par
\begin{figure*}
    \centering
    \includegraphics[width=\textwidth]{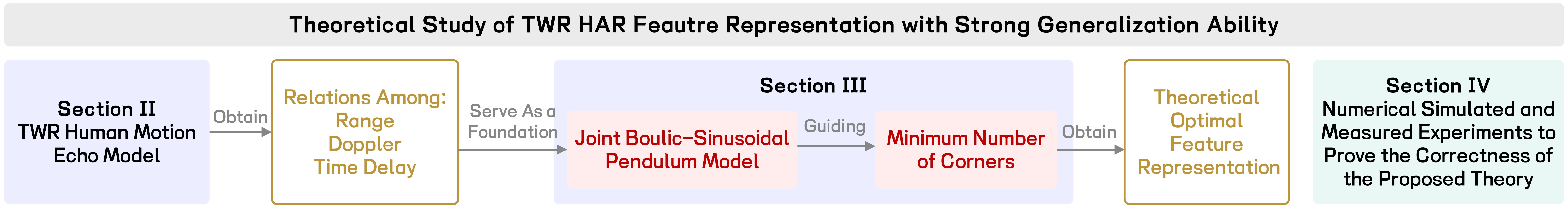}
    \caption{The logic of the construction of the proposed theoretical work and the structure of the paper.}
    \label{Overall Diagram}
    \vspace{-0.2cm}
\end{figure*}\par
Over the past decade, researchers have made many contributions in the area of signal processing for radar-based indoor human monitoring. HAR required spatial motion information over a period of time, and the recognition of different activities required complex micro-Doppler signature decoding process \cite{Existing Method Basis}. Zhang \emph{et al.} employed Doppler radar and extracted micro-Doppler signature based on the difference of micro-motions between the warhead and decoys for human gait imaging \cite{Zhang}. Gao \emph{et al.} also improved and verified these theories by simulations \cite{Gao}. Smith \emph{et al.} used dynamic time warping to study radar micro-Doppler signature \cite{Smith}, and highlighted the difficulties for classification. Fioranelli \emph{et al.} presented human micro-Doppler signature data gathered by a multi-static radar system to discriminate unarmed and potentially armed personnel walking along different trajectories \cite{Fioranelli}. A feature extraction method based on micro-Doppler signature was proposed by Du \emph{et al.} to categorize ground moving targets into three kinds, i.e., single walking person, double people walking, and a moving wheeled vehicle \cite{Du}. Chen \emph{et al.} introduced a series of statistical techniques for automatic hand, arm or body gesture recognition \cite{Chen}, and some other influential works for improvement \cite{Zeng, Fioranelli2}. Although these methods considered the principle of micro-Doppler signature, their extraction process was achieved by parametric estimation or image transformation, resulted in lower accuracy.\par
With the rapid development of AI, many deep-learning-based methods have been proposed for HAR. For example, Li \emph{et al.} elaborated neural network for HAR using 1D, 2D, and 3D radar echoes \cite{Li}. Singh \emph{et al.} proposed a framework for accurate HAR based on sparse and non-uniform point clouds \cite{Singh}. Cheng \emph{et al.} proposed a novel range profile sequence driven end-to-end method, which employed random cropping training for better model performance \cite{Cheng}. Ahmad \emph{et al.} proposed AEN to achieve refined micro-Doppler signature extraction at low noise levels \cite{AEN}. Jia \emph{et al.} proposed a radar image classification scheme using ResNet \cite{CNN1}, which aimed to achieve higher network depth and better recognition accuracy. In order to learn the long range timing information, Yang \emph{et al.} proposed an LSTM-based classifier \cite{LSTM} to train and inference the sliced radar images. Guo \emph{et al.} proposed to use PGM \cite{PGM} to achieve radar image feature extraction with certain topological modeling capabilities. Furthermore, a variety of methods based on feature representation learning were investigated \cite{TWR-MCAE}, achieving finer-grained micro-Doppler signature extraction \cite{TWR-WSN-CRF} or faster inference speeds \cite{TWR-FMSN}, respectively. These methods, although solving the problem of low accuracy, were often implemented by directly iterating the pseudo-RGB mapping of the radar RTM or DTM, which meant that the interpretability of the micro-Doppler signature extraction was poor \cite{Interpretability}. In addition, neural networks needed to be trained on large data sets, but the radar data acquisition of human activity were difficult and limited in number (In the vast majority of instances, the amount of data was much smaller than the number of parameters in the network), which meant that the final trained model was very prone to overfitting \cite{Overfitting}. As a result, it was difficult for these methods to have good generalization capability to different testers.\par
Relevant results demonstrated that the generalization performance of machine learning could be effectively improved through data processing, feature selection, regularization, cross-validation, or model tuning \cite{Classical Machine Learning, Feature Selection}. Constrained by the data and resource requirements of radar systems, feature selection and dimension reduction are the best choices to improve the generalization capability. Therefore, it is necessary to propose a micro-Doppler signature representation method based on mathematical-physical background that has a much smaller number of parameters but effectively characterizes the complete micro-Doppler information \cite{Hao Ling}.\par
In this paper, \textbf{TWR human activity feature can be represented by $\mathbf{30}$ key points for both range and Doppler profiles, which can effectively represent the complete micro-Doppler information while maximizing the generalization capability of indoor HAR across different testers.} The key points reflect the inflection points, stationary points, curve intersections, and intersections of the curves with the axes for each limb node on both range profiles and the Doppler profiles. These key points are defined in the rest of this paper as “Micro-Doppler Corner Feature”. Based on the works of Boulic and Thalmann \cite{Boulic Model}, in this paper, a kinematic model dedicated for feature dimension reduction analysis is presented, and the minimum number of corner points required to be able to represent the complete micro-Doppler information of various types of indoor human activities is calculated. Moreover, simulation and real-world experiments are conducted to demonstrate the effectiveness, efficiency, and generalization capability of the proposed micro-Doppler corner representation.\par
As shown in Fig. \ref{Overall Diagram}, the rest of the paper is organized as follows: Section II introduces the TWR signal model, followed by Section III, which introduces the proposed human motion model, and discusses the minimum number of corners. Section IV gives numerical simulated and measured experimental verification. Section V gives the conclusion. Details of a feasible feature extractor can be found in the Appendix.\par

\vspace{0.1cm}
\section{TWR-Based Human Motion Echo Model}
Human target is approximately equivalent to a superposition of different limb nodes as scattering centers. The distance and relative velocity of each node are reflected as the time delay and Doppler information in the echo \cite{Radar Echo Modeling}. Assuming the radar emits LFMCW signal. One CPI contains $M$ PRI. Then the time domain signal in the $m^\mathrm{th}$ PRI can be expressed as:

\vspace{-0.4cm}
\begin{equation}
\begin{gathered}S_{\mathrm{tx},m}(t)=A_{\mathrm{tx}}e^{j\left(2\pi\left(f_c(t-mT_s) +\frac{1}{2}\mu\left(t-mT_s\right)^2\right)+\varphi_\mathrm{tx}\right)}\\ mT_s\leq t\leq(m+1)T_s\end{gathered},
\end{equation}
where $A_{\mathrm{tx}}$ is the amplitude. $T_s$ is the PRI. $\mu=\frac{B}{T_s}$ is the chirp rate, $B$ is the bandwidth. $f_c$ is the carrier frequency. $\varphi_\mathrm{tx}$ is the initial phase.\par
Assume that the $N_i$ node of the body currently under consideration has a reflectivity of $\eta_{N_i}$. Introduce a two-way time delay $\tau_{N_i}(t) = \frac{2\xi_{N_i}}{c}$, where $c$ is the speed of light. Then, the time-domain echo signal of the radar can be expressed as:

\vspace{-0.4cm}
\begin{equation}
\begin{gathered}S_{\mathrm{rx},N_i,m}(t)=A_{\mathrm{rx},N_i}e^{j\left(2\pi\left(f_c t_{N_i} +\frac{1}{2}\mu t_{N_i}^2\right)+\varphi_\mathrm{rx}\right)}\\ t_{N_i}=t-mT_s-\tau_{N_i}(t),~mT_s\leq t\leq(m+1)T_s\end{gathered},
\end{equation}
where $A_{\mathrm{rx},N_i}=\eta_{N_i} A_{\mathrm{tx}}$, $\varphi_\mathrm{rx}=\varphi_\mathrm{tx}$. After mixing the echo with the transmitted signal, the following can be obtained:

\vspace{-0.2cm}
\begin{equation}
\begin{aligned}
S_{\mathrm{mix},N_i,m}(t)  &= S_{\mathrm{tx},m}(t) \cdot S_{\mathrm{rx},N_i,m}(t)\\&= A_{\mathrm{tx}}A_{\mathrm{rx},N_i}e^{jf_{\mathrm{mix},N_i}(t)},
\end{aligned}
\end{equation}
where,

\vspace{-0.3cm}
\begin{equation}
\label{Beat Signal}
\begin{aligned}
f_{\mathrm{mix},N_i}(t)&=2\pi f_c(t-mT_s) +\pi\mu(t-mT_s)^2\\&+\varphi_\mathrm{tx} +\varphi_\mathrm{rx}+2\pi f_c(t-mT_s-\tau_{N_i}(t))\\&+\pi\mu(t-mT_s-\tau_{N_i}(t))^2 \\
& = 4\pi f_c(t-mT_s)-2\pi f_c\tau_{N_i}(t)\\&+2\pi \mu (t-mT_s)^2+2\varphi_\mathrm{tx}\\
&-2\pi \mu (t-mT_s)\tau_{N_i}(t)+\pi \mu \tau^2_{N_i}(t)\\
mT_s&\leq t\leq(m+1)T_s
\end{aligned}.
\end{equation}\par
Thus, after low-pass filtering, the base-band echo of the $N_i$ node of the human target is obtained:

\vspace{-0.3cm}
\begin{equation}
\begin{aligned}
S_{b,N_i,m}(t) & = \frac{1}{2}A_{\mathrm{rx},N_i}A_{\mathrm{tx}} e^{jf_{\mathrm{mix,re},N_i}(t)} \\
& =A_{b,N_i}e^{j2\pi\left(f_c\tau_{N_i}(t)-\frac{1}{2}\mu\tau^2_{N_i}(t)+\mu(t-mT_s)\tau(t)\right)}  \\
& =A_{b,N_i} e^{j2\pi\left(f_c\tau_{N_i}(t)-\frac{1}{2}\mu\tau^2_{N_i}(t)+\mu(t-mT_s)\tau_{N_i}(t)\right)}\\A_{b,N_i}  &= \frac{1}{2}\eta_{N_i}A^2_{\mathrm{tx}},~mT \leq t\leq(m+1)T  
\end{aligned}.
\end{equation}\par
Summing radar echoes for different human limb nodes defined in TABLE \ref{Motion Constraints}, the complete time domain base-band radar echo is obtained:

\vspace{-0.3cm}
\begin{equation}
\begin{aligned}
S_{b,m}(t) & =\sum_{i=1}^{6}  S_{\mathrm{b},N_i,m}(t)+S_{b,\mathrm{Wall},m}+S_{b,\mathrm{Noise},m}\\
&=\sum_{i=1}^{6} A_{b,N_i} e^{j2\pi\left(f_c\tau_{N_i}(t)-\frac{1}{2}\mu\tau^2_{N_i}(t)+\mu(t-mT_s)\tau_{N_i}(t)\right)}\\ &+S_{b,\mathrm{Wall},m}+S_{b,\mathrm{Noise},m}
\end{aligned},
\label{Radar Echo}
\end{equation}
where $S_{b,\mathrm{Wall},m}$ is the echo component of the wall, and $S_{b,\mathrm{Noise},m}$ is the noise component.\par
\begin{table}
\begin{center}
\caption{Constraints of the Proposed Human Motion Model$^{*}$.\label{Motion Constraints}}
\vspace{-0.0cm}
\resizebox{0.48\textwidth}{!}{
\begin{tabular}{ccc}
\hline\hline 
\textbf{Name of Nodes} & \textbf{Initial Position (m)} & \textbf{Initial Velocity (m/s)}\\
\hline
Head ($N_1$) & $(x_1,y_1,h_1+0.15)$ & $(v_\mathrm{1x},v_\mathrm{1y},0)$\\
Torso ($N_2$) & $(x_1,y_1,\frac{h_1+h_2}{2})$ & $(v_\mathrm{1x},v_\mathrm{1y},0)$ \\
\hline
\textbf{Name of Nodes} & \multicolumn{2}{c}{\textbf{Constraints}$^{1}$} \\
\hline
Hands ($N_3~\&~N_4$) & \multicolumn{2}{c}{$\mid\mid \mathbf{Hand}-\mathbf{Torso}_1\mid\mid = l_1$}\\
Feet ($N_5~\&~N_6$) & \multicolumn{2}{c}{$\mid\mid \mathbf{Foot}-\mathbf{Torso}_2\mid\mid = l_2$}\\
\hline\hline
\end{tabular}
}
\end{center}
\footnotesize $^{*}$ The six points are determined by a combination of the key information in human motion, as well as the TWR wavelength with range resolution.\\
\footnotesize $^{1}$ $\textbf{Hand}$, $\textbf{Foot}$, $\textbf{Torso}_1$, and $\textbf{Torso}_2$ denote the coordinates of hand nodes, foot nodes, upper torso node, and lower torso node, respectively.\\
\vspace{-0.0cm}
\end{table}\par
\begin{table}
\begin{center}
\caption{Examples of Human Limb Nodes' Motion$^{*}$.}\label{Motion States Examples}
\vspace{-0.0cm}
\resizebox{0.48\textwidth}{!}{
\begin{tabular}{ccc}
\hline\hline 
\textbf{Serial Number} & \textbf{Motion States} & \textbf{Nodes Included}\\
\hline
\multicolumn{3}{c}{\textbf{Natural Walking}$^{1}$}\\
\hline
(1) & Acceleration-Free Motion & $N_1, N_2$\\
(2) & Sinusoidal Pendulum & $N_3,N_4,N_5,N_6$\\
(3) & Sudden Acceleration & $/$\\
\hline
\multicolumn{3}{c}{\textbf{In-situ Acceleration}$^{2}$}\\
\hline
(1) & Acceleration-Free Motion & $/$\\
(2) & Sinusoidal Pendulum & $/$\\
(3) & Sudden Acceleration & $N_1,N_2,N_3,N_4,N_5,N_6$\\
\hline
\multicolumn{3}{c}{\textbf{Combinations}$^{3}$}\\
\hline
(1) & Acceleration-Free Motion & $N_1,N_2$\\
(2) & Sinusoidal Pendulum & $N_3,N_4,N_5,N_6$\\
(3) & Sudden Acceleration & $N_1,N_2,N_3,N_4,N_5,N_6$\\
\hline\hline
\end{tabular}
}
\end{center}
\footnotesize $^{*}$ Two typical activities are considered. Other indoor human activities can be corresponded to each motion states in turn. "$/$" denotes that none of the nodes is associated with the current motion state. Definitions of limb nodes $N_1 \sim N_6$ are consistent in TABLE \ref{Motion Constraints}.\\
\footnotesize $^{1}$ Activities that can be approximated to be classified as natural walking include walking, kicking, punching, and rotation. The design idea is that the motion state of each limb node in these activities can be seen in a state of periodic oscillation.\\
\footnotesize $^{2}$ Activities that can be approximated to be classified as in-situ acceleration include grabbing, standing up, and sitting down. The design idea is that the motion state of each limb node in these activities can be seen as a falling process of first acceleration and then deceleration.\\
\footnotesize $^{3}$ Activities that can be approximated to be classified as the combinations include walking to sitting, sitting to walking, walking to falling, and falling to walking. The design idea is that the limb nodes of the human body swing periodically around a certain center, and at the same time there are acceleration and deceleration processes in that center.\\
\vspace{-0.2cm}
\end{table}\par

\section{Optimal Corner Representation of Micro-Doppler Signature}
In this section, we first develop the joint Boulic-sinusoidal pendulum model. Then, we discuss the minimum required number of corners to maximize feature dimension reduction while preserving human motion information.\par
\begin{figure*}
    \centering
    \includegraphics[width=\textwidth]{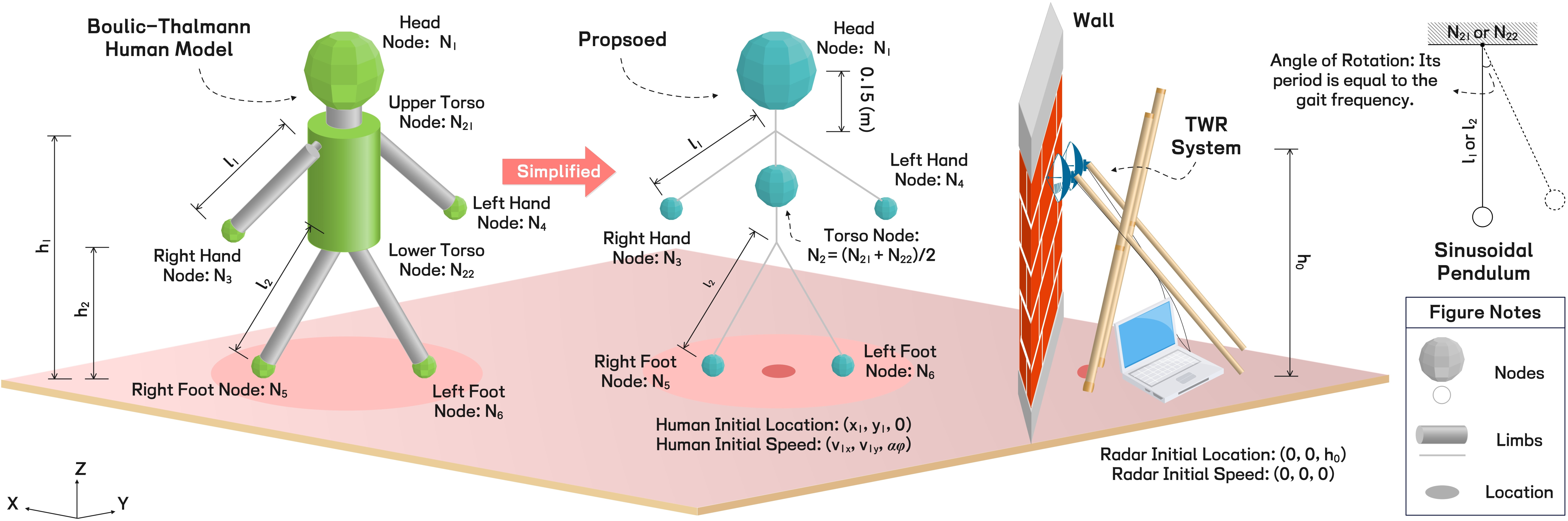}
    \caption{Schematic diagram of the proposed joint Boulic-sinusoidal pendulum model.}
    \label{Boulic Motion Model}
    \vspace{-0.0cm}
\end{figure*}\par
\subsection{Joint Boulic-Sinusoidal Pendulum Model}
As shown in Fig. \ref{Boulic Motion Model}, the proposed joint Boulic-sinusoidal pendulum model consists of two stages. The first is human structure model, which is the simplified version of the Boulic-Thalmann limb-node model \cite{Boulic Model}. The second is gait pattern model, which utilizes an approximation of the pendulum model with a sinusoidal period \cite{Pendulum Model}. Considering the range and Doppler resolution of UWB radars \cite{Guolong Cui}, human structure model contains only seven nodes can effectively characterize the motion information: the head, the upper edge of the torso, the lower edge of the torso, both hands, and both feet. Note that the torso is designed as an individual node using the center-of-scattering method \cite{Center of Scattering}. Therefore, the human model proposed in this paper only contains six limb nodes in total.\par
Assume that the radar observation location is fixed at $(0,0,h_0)$, the initial position of the human body is $(x_1,y_1,0)$, the arm length is $l_1$, and the leg length is $l_2$. The heights of the upper and lower edges of the torso from the ground are $h_1$ and $h_2$, respectively. The initial speed of the body is $(v_\mathrm{1x},v_\mathrm{1y},\alpha\varphi)$, where $\alpha$ is the magnitude of human micro-undulation motion in Z-direction, $\varphi$ is the gait frequency. The resulting motion constraints for the six nodes are shown in TABLE \ref{Motion Constraints}. For common human indoor activities \cite{Amin1}, the micro-Doppler signature can be regarded as: (1) Natural Walking, (2) In-situ Acceleration, or (3) Their combinations shown in TABLE \ref{Motion States Examples}. Therefore, in the following, the mathematical models of range and velocity curves for different nodes under (1) and (2) are derived, respectively.\par
\quad \par
\noindent \textbf{1. Natual Walking}\par
\textbf{1.1. Head Node $N_1$}\par
At moment $t$, the distance between the head node $N_1$ and radar is:

\vspace{-0.2cm}
\begin{equation}
\begin{aligned}
R_\mathrm{He} & = \left(R_1^{\prime2}+2(x_1 v_\mathrm{1x}+y_1 v_\mathrm{1y})t+v_1^2 t^2 \right. \\& \left. -h_1^2+(h_1-h_0+\omega(t)+0.15)^2 \right)^{\frac{1}{2}},
\end{aligned}
\end{equation}
where:

\vspace{-0.2cm}
\begin{equation}
R_1^{\prime} = \sqrt{x_1^2+y_1^2+h_1^2},\quad v_1 = \sqrt{v_\mathrm{1x}^2+y_\mathrm{1y}^2},
\end{equation}
and:

\vspace{-0.2cm}
\begin{equation}
\omega(t)=\alpha \sin(\varphi t),    
\end{equation}
is the micro-undulation process during human motion, $h_1-h_0+0.15 \gg \alpha$. The effect on velocity is then superimposed component in the z-axis that:

\vspace{-0.2cm}
\begin{equation}
\omega^{\prime}(t)=\alpha \varphi \cos(\varphi t).
\end{equation}\par
The distance $R_\mathrm{He}$ can be rewritten as:  

\vspace{-0.2cm}
\begin{equation}
R_\mathrm{He} \approx \left({R_1^2+2(x_1 v_\mathrm{1x}+y_1 v_\mathrm{1y})t+v_1^2 t^2}\right)^{\frac{1}{2}},
\end{equation}
where:

\vspace{-0.2cm}
\begin{equation}
R_1 = \sqrt{x_1^2+y_1^2+(h_1-h_0+0.15)^2}.   
\end{equation}\par
In practice, $R_{\mathrm{He},\mathrm{real}}=R_\mathrm{He}$ for free space, and:

\vspace{-0.2cm}
\begin{equation}
R_{\mathrm{He},\mathrm{real},\mathrm{TWR}}\approx R_\mathrm{He}+d(\sqrt{\epsilon_r}-1),
\end{equation}
for through-the-wall detection, where $d,\epsilon_r$ are the thickness and relative permittivity of the wall, respectively.\par
Define one-way propagation distance as $\xi_{N_1}$. Then, 

\vspace{-0.2cm}
\begin{equation}
\xi_{N_1}^2 = v_1^2 t^2+2(x_1 v_\mathrm{1x}+y_1 v_\mathrm{1y})t+R_1^2,
\end{equation}
which is a quadratic function with respect to $t$. Thus, we only need three different points to uniquely determine the analytic form of the function. In general, the two points intersecting the front and back edges of the observed time window and the extreme points are used:\par
(1) When $t=0$, we get:

\vspace{-0.2cm}
\begin{equation}
\xi_{N_1}^2=R_1^2.
\end{equation}\par
(2) When $t=T$, we get:

\vspace{-0.2cm}
\begin{equation}
\xi_{N_1}^2 = v_1^2 T^2+2(x_1 v_\mathrm{1x}+y_1 v_\mathrm{1y})T+R_1^2,
\end{equation}
where $T$ is the sample duration.\par
(3) When $t=-\frac{v_\mathrm{1x}}{v_1^2}x_1-\frac{v_\mathrm{1y}}{v_1^2}y_1 \in [0,T]$, we get:

\vspace{-0.2cm}
\begin{equation}
\xi_{N_1}^2=R_1^2-(\frac{v_\mathrm{1x}}{v_1}x_1+\frac{v_\mathrm{1y}}{v_1}y_1)^2.
\end{equation}\par
(4) When $t=-\frac{v_\mathrm{1x}}{v_1^2}x_1-\frac{v_\mathrm{1y}}{v_1^2}y_1 \in (-\infty,0)\cup (T,\infty)$, take any $t=t_1 \in [0,T]$, we get: 

\vspace{-0.2cm}
\begin{equation}
\xi_{N_1}^2 = v_1^2 t_1^2+2(x_1 v_\mathrm{1x}+y_1 v_\mathrm{1y})t_1+R_1^2.
\end{equation}\par
In addition, the squared velocity curve of the head node is:

\vspace{-0.2cm}
\begin{equation}
\chi_{N_1}^2=V_\mathrm{He}^2 = v_\mathrm{1x}^2+v_\mathrm{1y}^2+\omega^{\prime 2}(t).
\end{equation}\par
Consider that the derivative of the micro-undulation movement is extremely small, we get:

\vspace{-0.2cm}
\begin{equation}
\Rightarrow \quad \chi_{N_1}^2 \approx v_\mathrm{1x}^2+v_\mathrm{1y}^2,  
\end{equation}
in which the principle component is constant, so only one point is enough.\par
\quad \par
\textbf{1.2. Torso Node $N_2$}\par
Similarly, for the torso node, the analysis process is exactly the same as for the head node.\par
Here, define $R_2$ to replace $R_1$:

\vspace{-0.2cm}
\begin{equation}
R_2 = \sqrt{x_1^2+y_1^2+(\frac{h_1+h_2}{2}-h_0)^2}.
\end{equation}\par
Thus, the squared one-way propagation distance $\xi^2_{N_2}$ is:

\vspace{-0.2cm}
\begin{equation}
\xi_{N_2}^2 = v_1^2 t^2+2(x_1 v_\mathrm{1x}+y_1 v_\mathrm{1y})t+R_2^2,
\end{equation}
which is also a quadratic function with respect to $t$. The effect of wall refraction on the distance curve is still the addition of a $d(\sqrt{\epsilon_r}-1)$ term, which does not affect the order of the curve. (It will also not to be considered subsequently.) Thus, three different points are enough. In addition, the squared velocity curve of torso node $\chi^2_{N_2}$ is the same as the head node:

\vspace{-0.2cm}
\begin{equation}
\begin{aligned}
\chi_{N_2}^2&=\chi_{N_1}^2=V_\mathrm{He}^2 \\ &= v_\mathrm{1x}^2+v_\mathrm{1y}^2+\omega^{\prime 2}(t)
\end{aligned},
\end{equation}

\vspace{-0.2cm}
\begin{equation}
\Rightarrow \quad \chi_{N_2}^2 \approx v_\mathrm{1x}^2+v_\mathrm{1y}^2. 
\end{equation}\par
According to the above analysis, only one point is needed to reconstruct the original function.\par
\quad \par
\textbf{1.3. Hand Nodes $N_3~\&~N_4$}\par
Assume that the human arm is a radial rigid body. In the sinusoidal pendulum model, the maximum angle between the swinging arm and the vertical direction is $\theta_1$. Similarly its squared one-way propagation distance curve is:

\vspace{-0.3cm}
\begin{equation}
\label{Hand, Distance, Walking}
\begin{aligned}
\xi_{N_3}^2 &=  R_3^2+2x_1 v_\mathrm{1x} t+2y_1 v_\mathrm{1y} t+v_1^2 t^2\\
& +\frac{2l_1}{v_1}l_1\sin(\theta_1 \sin(\varphi t))(x_1 v_\mathrm{1x}+y_1 v_\mathrm{1y}+v_1^2 t)\\
& -2h_1 l_1 \cos(\theta_1 \sin(\varphi t))
\end{aligned},
\end{equation}
where:

\vspace{-0.2cm}
\begin{equation}
R_3  = \sqrt{x_1^2+y_1^2+h_1^2+l_1^2}.
\end{equation}\par
The term $\sin(\theta_1 \sin(\varphi t))$ is an odd, periodic, and bounded function, with a maximum value of $\sin(\theta_1)$, which can be uniquely determined by six points. Thus, in order to reconstruct the information in $\xi_{N_3}^2$, it is necessary to identify six points:\par
(1) When $t=0$, we get:

\vspace{-0.2cm}
\begin{equation}
\xi_{N_3}^2 = R_3^2-2h_1 l_1 = x_1^2+y_1^2+(h_1-l_1)^2
\end{equation}\par
(2) When $t=T$, we get:

\vspace{-0.2cm}
\begin{equation}
\begin{aligned}
\xi_{N_3}^2 &= R_3^2+2x_1 v_\mathrm{1x} T+2y_1 v_\mathrm{1y} T+v_1^2 T^2\\&+\frac{2l_1}{v_1}l_1\sin(\theta_1 \sin(\varphi T))(x_1 v_\mathrm{1x}+y_1 v_\mathrm{1y}+v_1^2 T)\\&-2h_1 l_1 \cos(\theta_1 \sin(\varphi T))    
\end{aligned}.
\end{equation}\par
(3) When $t \in (0,T)$, considering the following derivation that characterizes the monotonicity of the curve:

\vspace{-0.2cm}
\begin{equation}
\label{Hand, Distance, Walking, Derivation}
\begin{aligned}
\frac{\partial \xi_{N_3}^2}{\partial t} &= 2R_4(t)\\&+2R_4(t)\frac{l_1 \theta_1 \varphi}{v_1}\cos(\varphi t)\cos(\theta_1 \sin(\varphi t))\\
&+2l_1\sin(\theta_1 \sin(\varphi t))(v_1+h_1\theta_1 \varphi \cos(\varphi t))
\end{aligned},
\end{equation}
in which:

\vspace{-0.2cm}
\begin{equation}
R_4(t) = x_1 v_\mathrm{1x}+y_1 v_\mathrm{1y}+v_1^2 t,
\end{equation}
is not a constant function. If the zeros of Eq. (\ref{Hand, Distance, Walking, Derivation}) in $(0,T)$ are greater than or equal to $4$, then the corresponding $t_1 \sim t_4$ in $\frac{\partial \xi_{N_3}^2}{\partial t_i}\mid_{i=1\sim 4}=0$ are the rest four points we need. If the zeros of Eq. (\ref{Hand, Distance, Walking, Derivation}) in $(0,T)$ are less than $4$, then we continue by choosing the second-order derivative zeros as the desired $t$ value. In the vast majority of cases, human gait habits usually satisfy:

\vspace{-0.2cm}
\begin{equation}
\frac{T\varphi}{\pi}\geq 2,
\end{equation}
which means:

\vspace{-0.2cm}
\begin{equation}
T \geq \frac{2 \pi}{\min(\varphi)}.
\end{equation}\par
It is usually only necessary to use at most the second-order derivative zeros to pick the remaining four $t$ values.\par
The squared velocity curve of the hand node $N_3$ is constructed using the vertex velocity of a sinusoidal pendulum:

\vspace{-0.2cm}
\begin{equation}
\begin{aligned}
\chi_{N_3}^2&=V_\mathrm{Ha}^2\\ &= v_1^2-2l_1v_1\theta_1\varphi \cos(\varphi t)\cos(\theta_1 \sin(\varphi t))\\&+l_1^2\theta_1^2\varphi^2\cos^2(\varphi t)
\end{aligned}.
\end{equation}\par
\begin{table*}
\begin{center}
\caption{Calculation of Minimum Number of Corner Points in Range and T-F Profiles$^{*}$.\label{Minimum Corner Points}}
\vspace{-0.4cm}
\resizebox{\textwidth}{!}{
\begin{tabular}{cccccc}
\hline\hline 
\textbf{Motion} & \textbf{Nodes} & \textbf{Curve Equations on $\mathbf{R^2TM}$} & \textbf{Curve Equations on $\mathbf{D^2TM}$}& \textbf{MNCP ($\mathbf{R^2TM}$)}& \textbf{MNCP ($\mathbf{D^2TM}$)}\\
\hline
\multirow{15}{*}{Mot. $1^{1}$} & $N_1$ & $R_{N_1,t}^2=a_{N_1,0}+a_{N_1,1}t+a_{N_1,2}t^2$& $D_{N_1,t}^2=b_{N_1,0}$& $3$ & $1$\\
& $N_2$ & $R_{N_2,t}^2=a_{N_2,0}+a_{N_2,1}t+a_{N_2,2}t^2$ & $D_{N_2,t}^2=b_{N_2,0}$& $3$ & $1$\\
& $N_3$ & $\begin{gathered} R_{N_3,t}^2=(a_{N_3,0}+a_{N_3,1}\sin(a_{N_3,2}\sin(a_{N_3,3}t)))\\ \cdot(a_{N_3,4}+a_{N_3,5}t+a_{N_3,6}t^2)\\+a_{N_3,7}\cos(a_{N_3,8}\sin(a_{N_3,9}t))\end{gathered}$ & $\begin{gathered} D_{N_3,t}^2 =b_{N_3,0}\\+ b_{N_3,1}\cos^2( b_{N_3,2}t)\\+b_{N_3,3}\cos(b_{N_3,4}t) \cos(b_{N_3,5}\sin(b_{N_3,6}t))\end{gathered}$ & $6$ & $5$\\
& $N_4$ & $\begin{gathered} R_{N_4,t}^2=(a_{N_4,0}+a_{N_4,1}\sin(a_{N_4,2}\sin(a_{N_4,3}{t^{\prime}})))\\ \cdot(a_{N_4,4}+a_{N_4,5}{t^{\prime}}+a_{N_4,6}{t^{\prime}}^2)\\+a_{N_4,7}\cos(a_{N_4,8}\sin(a_{N_4,9}{t^{\prime}}))\end{gathered}$ & $\begin{gathered} D_{N_4,t}^2 =b_{N_4,0}\\+ b_{N_4,1}\cos^2( b_{N_4,2}{t^{\prime}})\\+b_{N_4,3}\cos(b_{N_4,4}{t^{\prime}}) \cos(b_{N_4,5}\sin(b_{N_4,6}{t^{\prime}}))\end{gathered}$ & $6$ & $5$\\
& $N_5$ & $\begin{gathered} R_{N_5,t}^2=(a_{N_5,0}+a_{N_5,1}\sin(a_{N_5,2}\sin(a_{N_5,3}{t^{\prime}})))\\ \cdot(a_{N_5,4}+a_{N_5,5}{t^{\prime}}+a_{N_5,6}{t^{\prime}}^2)\\+a_{N_5,7}\cos(a_{N_5,8}\sin(a_{N_5,9}{t^{\prime}}))\end{gathered}$ & $\begin{gathered} D_{N_5,t}^2 =b_{N_5,0}\\+ b_{N_5,1}\cos^2( b_{N_5,2}{t^{\prime}})\\+b_{N_5,3}\cos(b_{N_5,4}{t^{\prime}}) \cos(b_{N_5,5}\sin(b_{N_5,6}{t^{\prime}}))\end{gathered}$ & $6$ & $5$\\
& $N_6$ & $\begin{gathered} R_{N_6,t}^2=(a_{N_6,0}+a_{N_6,1}\sin(a_{N_6,2}\sin(a_{N_6,3}t)))\\ \cdot(a_{N_6,4}+a_{N_6,5}t+a_{N_6,6}t^2)\\+a_{N_6,7}\cos(a_{N_6,8}\sin(a_{N_6,9}t))\end{gathered}$ & $\begin{gathered} D_{N_6,t}^2 =b_{N_6,0}\\+ b_{N_6,1}\cos^2( b_{N_6,2}t)\\+b_{N_6,3}\cos(b_{N_6,4}t) \cos(b_{N_6,5}\sin(b_{N_6,6}t))\end{gathered}$ & $6$ & $5$\\
& $\text{Total}$ & $R_t^2 = \sum_{i=1}^{6}R_{N_i,t}^2$ & $D_t^2 = \sum_{i=1}^{6}D_{N_i,t}^2$ & $30$ & $22$\\
\hline
\multirow{14}{*}{Mot. $2^{1}$}& $N_1$ & $\begin{gathered} R_{N_1,t}^2=a^{\prime}_{N_1,0}+a^{\prime}_{N_1,1}\sin(a^{\prime}_{N_1,2}t^{\prime\prime})\\+a^{\prime}_{N_1,3}\cos(a^{\prime}_{N_1,4}t^{\prime\prime}) \end{gathered}$ & $D_{N_1,t}^2 = b^{\prime}_{N_1,0}+b^{\prime}_{N_1,1}\cos(b^{\prime}_{N_1,2}t^{\prime\prime})$ & $5$ & $5$ \\
& $N_2$ & $\begin{gathered} R_{N_2,t}^2=a^{\prime}_{N_2,0}+a^{\prime}_{N_2,1}\sin(a^{\prime}_{N_2,2}t^{\prime\prime})\\+a^{\prime}_{N_2,3}\cos(a^{\prime}_{N_2,4}t^{\prime\prime}) \end{gathered}$ & $D_{N_2,t}^2 = b^{\prime}_{N_2,0}+b^{\prime}_{N_2,1}\cos(b^{\prime}_{N_2,2}t^{\prime\prime})$ & $5$ & $5$ \\
& $N_3$ & $\begin{gathered} R_{N_3,t}^2=a^{\prime}_{N_3,0}+a^{\prime}_{N_3,1}\sin(a^{\prime}_{N_3,2}t^{\prime\prime})\\+a^{\prime}_{N_3,3}\cos(a^{\prime}_{N_3,4}t^{\prime\prime}) \end{gathered}$ & $D_{N_3,t}^2 = b^{\prime}_{N_3,0}+b^{\prime}_{N_3,1}\cos(b^{\prime}_{N_3,2}t^{\prime\prime})$ & $5$ & $5$ \\
& $N_4$ & $\begin{gathered} R_{N_4,t}^2=a^{\prime}_{N_4,0}+a^{\prime}_{N_4,1}\sin(a^{\prime}_{N_4,2}t^{\prime\prime})\\+a^{\prime}_{N_4,3}\cos(a^{\prime}_{N_4,4}t^{\prime\prime}) \end{gathered}$ & $D_{N_4,t}^2 = b^{\prime}_{N_4,0}+b^{\prime}_{N_4,1}\cos(b^{\prime}_{N_4,2}t^{\prime\prime})$ & $5$ & $5$ \\
& $N_5$ & $\begin{gathered} R_{N_5,t}^2=a^{\prime}_{N_5,0}+a^{\prime}_{N_5,1}\sin(a^{\prime}_{N_5,2}t^{\prime\prime})\\+a^{\prime}_{N_5,3}\cos(a^{\prime}_{N_5,4}t^{\prime\prime}) \end{gathered}$ & $D_{N_5,t}^2 = b^{\prime}_{N_5,0}+b^{\prime}_{N_5,1}\cos(b^{\prime}_{N_5,2}t^{\prime\prime})$ & $5$ & $5$ \\
& $N_6$ & $\begin{gathered} R_{N_6,t}^2=a^{\prime}_{N_6,0}+a^{\prime}_{N_6,1}\sin(a^{\prime}_{N_6,2}t^{\prime\prime})\\+a^{\prime}_{N_6,3}\cos(a^{\prime}_{N_6,4}t^{\prime\prime}) \end{gathered}$ & $D_{N_6,t}^2 = b^{\prime}_{N_6,0}+b^{\prime}_{N_6,1}\cos(b^{\prime}_{N_6,2}t^{\prime\prime})$ & $5$ & $5$ \\
& $\text{Total}$ & $R_t^2 = \sum_{i=1}^{6}R_{N_i,t}^2$ & $D_t^2 = \sum_{i=1}^{6}D_{N_i,t}^2$ & $30$ & $30$\\
\hline\hline
\end{tabular}
}
\end{center}
\footnotesize $^{*}$ All the modulus in this table, including $a_{N_i,j},b_{N_i,j},a^{\prime}_{N_i,j},b^{\prime}_{N_i,j},i\in \mathcal{N},j \in \mathcal{N} $ are constants related to the human body posture and gait parameters that are independent of time $t$. $t^{\prime}=t-\pi$, $t^{\prime\prime}=t-t_0$, where $t_0$ is a quarter cycle of the previously defined motion without position change. $N_1\sim N_6$ are consistent with the definition in Fig. \ref{Boulic Motion Model}. $R_{N_i,t}^2,D_{N_i,t}^2,i=1\sim 6$ represent the vertical axis in $\mathbf{R^2TM}$ and $\mathbf{D^2TM}$, respectively.\\
\footnotesize $^{1}$ Mot. $1$ and Mot. $2$ denote the natural walking and in-situ acceleration discussed in section II, respectively.\\
\vspace{-0.2cm}
\end{table*}\par
According to the above analysis, five different points are needed to reconstruct the original function:\par
(1) When $t=0$, we get:

\vspace{-0.2cm}
\begin{equation}
V_\mathrm{Ha}^2 = (v_1-l_1 \theta_1 \varphi)^2.
\end{equation}\par
(2) When $t=T$, we get:

\vspace{-0.2cm}
\begin{equation}
\begin{aligned}
V_\mathrm{Ha}^2 &= v_1^2-2l_1v_1\theta_1\varphi \cos(\varphi T)\cos(\theta_1 \sin(\varphi T))\\&+l_1^2\theta_1^2\varphi^2\cos^2(\varphi T)
\end{aligned}.
\end{equation}\par
(3) When $t \in (0,T)$, consider the following derivation:

\vspace{-0.2cm}
\begin{equation}
\begin{aligned}
\frac{\partial V_\mathrm{Ha}^2}{\partial t} &= 2l_1v_1\theta_1\varphi^2 \sin(\varphi t) \cos(\theta_1 \sin(\varphi t))\\ & +2l_1v_1\theta_1^2\varphi^2\cos^2(\varphi t) \sin(\theta_1 \sin(\varphi t))\\ &-2l_1^2\theta_1^2\varphi^3 \sin(\varphi t)\cos(\varphi t)
\end{aligned}.
\end{equation}\par
We take $t_i$ in $\frac{\partial V_\mathrm{Ha}^2}{\partial t_i}\mid_{i=1\sim 3}=0$ for the rest selection of $t$ value. If there are not enough extreme points, the remaining $t_i$ are selected by second derivative zeros $\frac{\partial^2 V_\mathrm{Ha}^2}{\partial t_i^2}=0$.\par 
For hand node $N_4$, the squared distance curve can be expressed as the transformation of $\xi^2_{N_3}$ by $\pi$ phase:

\vspace{-0.3cm}
\begin{equation}
\begin{aligned}
\xi_{N_4}^2 &=  R_3^2+2x_1 v_\mathrm{1x} t+2y_1 v_\mathrm{1y} t+v_1^2 t^2\\
& +\frac{2l_1}{v_1}l_1\sin(\theta_1 \sin(\varphi t)+\pi)(x_1 v_\mathrm{1x}+y_1 v_\mathrm{1y}+v_1^2 t)\\
& -2h_1 l_1 \cos(\theta_1 \sin(\varphi t)+\pi)
\end{aligned}.
\end{equation}\par
The squared velocity curve of the hand node $N_4$ is :

\vspace{-0.2cm}
\begin{equation}
\begin{aligned}
\chi_{N_4}^2 &= v_1^2-2l_1v_1\theta_1\varphi \cos(\varphi t)\cos(\theta_1 \sin(\varphi t)+\pi)\\&+l_1^2\theta_1^2\varphi^2\cos^2(\varphi t)
\end{aligned}.
\end{equation}\par
Therefore, six and five different points are needed to reconstruct the squared distance and velocity curves, respectively.\par
\begin{algorithm}[htbp]
\DontPrintSemicolon
  \KwInput{$\mathbf{I_r}=\mathbf{RTM}$, $\mathbf{I_d}=\mathbf{DTM}$}
  \KwOutput{Range and T-F profiles $\mathbf{R^{2}TM},\mathbf{D^{2}TM}$ after vertical axis squaring.}
  $[s,l] = \mathrm{size}(\mathbf{I}_r^{\top})$, $[p,q] = \mathrm{size}(\mathbf{I}_d^{\top})$, $i=j=k=0$;\;
  \tcc{$\mathrm{size}()$ denotes the dimension of the matrix.}
  $\mathbf{I}_{r,2} = \mathrm{zeros}(s,l^2)$;\;
  \tcc{$\mathrm{zeros}()$ is to create an all-zero matrix.}
  \If{$q \equiv 2~(\mathrm{mod}~2)$}{$\mathbf{I}_{d,1} = \mathbf{I}_{d,2} = \mathrm{zeros}(p,q^2/4)$;\;\Else{$\mathbf{I}_{d,1} = \mathbf{I}_{d,2} = \mathrm{zeros}(p,(q+1)^2/4)$;}}
  \For{$i = 1:s$}{
     \For{$j = 1:l$}{
        \For{$k = 1:(2j-1)$}{
            $\mathbf{I}_{r,2}(i,(j-1)^2+k) =\mathbf{I}_r^\top(i,j)$;\;
  }
  }
  }
  \For{$i = 1:p$}{
     \For{$j = 1:\lfloor\frac{q}{2}\rfloor+1$}{
        \For{$k = 1:(2j-1)$}{
            $\mathbf{I}_{d,1}(i,(j-1)^2+k) =\mathbf{I}_d^\top(i,[q/2]-j)$;\;
            $\mathbf{I}_{d,2}(i,(j-1)^2+k) =\mathbf{I}_d^\top(i,[q/2]+j)$;\;
  }
  }
  }
  $\mathbf{R^2TM} = \mathrm{Norm}(\mathbf{I}_{r,2}^\top)$;\;
  $\mathbf{D^2TM} = \mathrm{Norm}(\mathrm{Con}(\mathrm{flip}(\mathbf{I}_{d,1}),\mathbf{I}_{d,2})^\top)$;\;
  \tcc{$\mathrm{Norm}()$ is the normalization function, $\mathrm{flip}()$ means flip the matrix vertically, and $\mathrm{Con}()$ means the concatenation of two matrices.}
\caption{Generation of Range and T-F Profiles}
\label{Squared in Vertical Axis}
\end{algorithm}\par
\quad \par
\textbf{1.4. Foot Nodes $N_5~\&~N_6$}\par
The analysis process for the foot nodes is exactly the same as the hand nodes, and only the definition of some parameters needs to be modified. For example, replace $l_1$, $h_1$, $\theta_1$ with $l_2$, $h_2$, $\theta_2$, respectively. It can be concluded: 

\vspace{-0.3cm}
\begin{equation}
\begin{aligned}
\xi_{N_5}^2 &= R_5^2+2x_1 v_\mathrm{1x} t+2y_1 v_\mathrm{1y} t+v_1^2 t^2\\
& +\frac{2l_2}{v_1}l_2\sin(\theta_2 \sin(\varphi t))(x_1 v_\mathrm{1x}+y_1 v_\mathrm{1y}+v_1^2 t)\\
& -2h_2 l_2 \cos(\theta_2 \sin(\varphi t))
\end{aligned},
\end{equation}

\vspace{-0.2cm}
\begin{equation}
\begin{aligned}
\xi_{N_6}^2 &= R_5^2+2x_1 v_\mathrm{1x} t+2y_1 v_\mathrm{1y} t+v_1^2 t^2\\
& +\frac{2l_2}{v_1}l_2\sin(\theta_2 \sin(\varphi t)+\pi)(x_1 v_\mathrm{1x}+y_1 v_\mathrm{1y}+v_1^2 t)\\
& -2h_2 l_2 \cos(\theta_2 \sin(\varphi t)+\pi)
\end{aligned},
\end{equation}
where:

\vspace{-0.2cm}
\begin{equation}
R_5=\sqrt{x_1^2+y_1^2+h_2^2+l_2^2},
\end{equation}
and:

\vspace{-0.2cm}
\begin{equation}
\begin{aligned}
\chi_{N_5}^2 &= v_1^2-2l_2v_1\theta_2\varphi \cos(\varphi t)\cos(\theta_2 \sin(\varphi t))\\&+l_2^2\theta_2^2\varphi^2\cos^2(\varphi t)
\end{aligned},
\end{equation}

\vspace{-0.2cm}
\begin{equation}
\begin{aligned}
\chi_{N_6}^2 &= v_1^2-2l_2v_1\theta_2\varphi \cos(\varphi t)\cos(\theta_2 \sin(\varphi t)+\pi)\\&+l_2^2\theta_2^2\varphi^2\cos^2(\varphi t)
\end{aligned}.
\end{equation}\par
In total, six and five different points are needed to reconstruct the squared distance and velocity curves, respectively.\par
\quad \par
\noindent \textbf{2. In-situ Acceleration}\par
The body of the human can be modeled solely by upper torso node located at height $h_1$. Then: 

\vspace{-0.2cm}
\begin{equation}
h_1= \frac{1}{2}\left(\Delta h_1 \sin(t-t_0)+h_1-\frac{1}{2}\Delta h_1\right),
\end{equation}
where $t_0$ is the half period of the in-situ acceleration motion, $T=4t_0$, $\Delta h_1$ is the maximum height variation for the upper torso node. Thus, the squared one-way propagation distance curve $\xi^2$ at this moment can be rewritten as:

\vspace{-0.2cm}
\begin{equation}
\label{Torso, Distance, Squatting}
\begin{aligned}
\xi^2 &= R_6^2+\frac{1}{8}\Delta h_1^2 +\Delta h_1 R_6^{\prime}\sin\frac{\pi}{2t_0}(t-t_0)\\&-\frac{1}{8}\Delta h_1^2 \cos\frac{\pi}{t_0}(t-t_0)
\end{aligned}
\end{equation}
where:

\vspace{-0.2cm}
\begin{equation}
R_6^{\prime} = h_1-h_0-\frac{1}{2}\Delta h_1,
\end{equation}

\vspace{-0.2cm}
\begin{equation}
R_6 = \sqrt{x_1^2+y_1^2+R_6^{\prime 2}}.
\end{equation}\par
The orders of four parts in Eq. (\ref{Torso, Distance, Squatting}) are $0,0,5,5$, respectively.\par
Furthermore, the squared velocity curve of the upper torso node is:

\vspace{-0.2cm}
\begin{equation}
\label{Torso, Velocity, Squatting}
V_\mathrm{UT}^2 = \frac{\pi}{32t_0^2}\Delta h_1^2\left(1+\cos\frac{\pi}{t_0}(t-t_0)\right),
\end{equation}
where the order of Eq. (\ref{Torso, Velocity, Squatting}) is also $5$. The resulting squared distance curves $\xi^2_i,i=N_1 \sim N_6$ and squared velocity curves $\chi^2_i,i=N_1 \sim N_6$ are obtained from $\xi^2,V^2_\mathrm{UT}$ after stretching transformation in vertical axis, respectively. Therefore, $5$ different points are needed to reconstruct both the squared distance and velocity curves, respectively.\par
\begin{figure}
    \centering
    \includegraphics[width=0.48\textwidth]{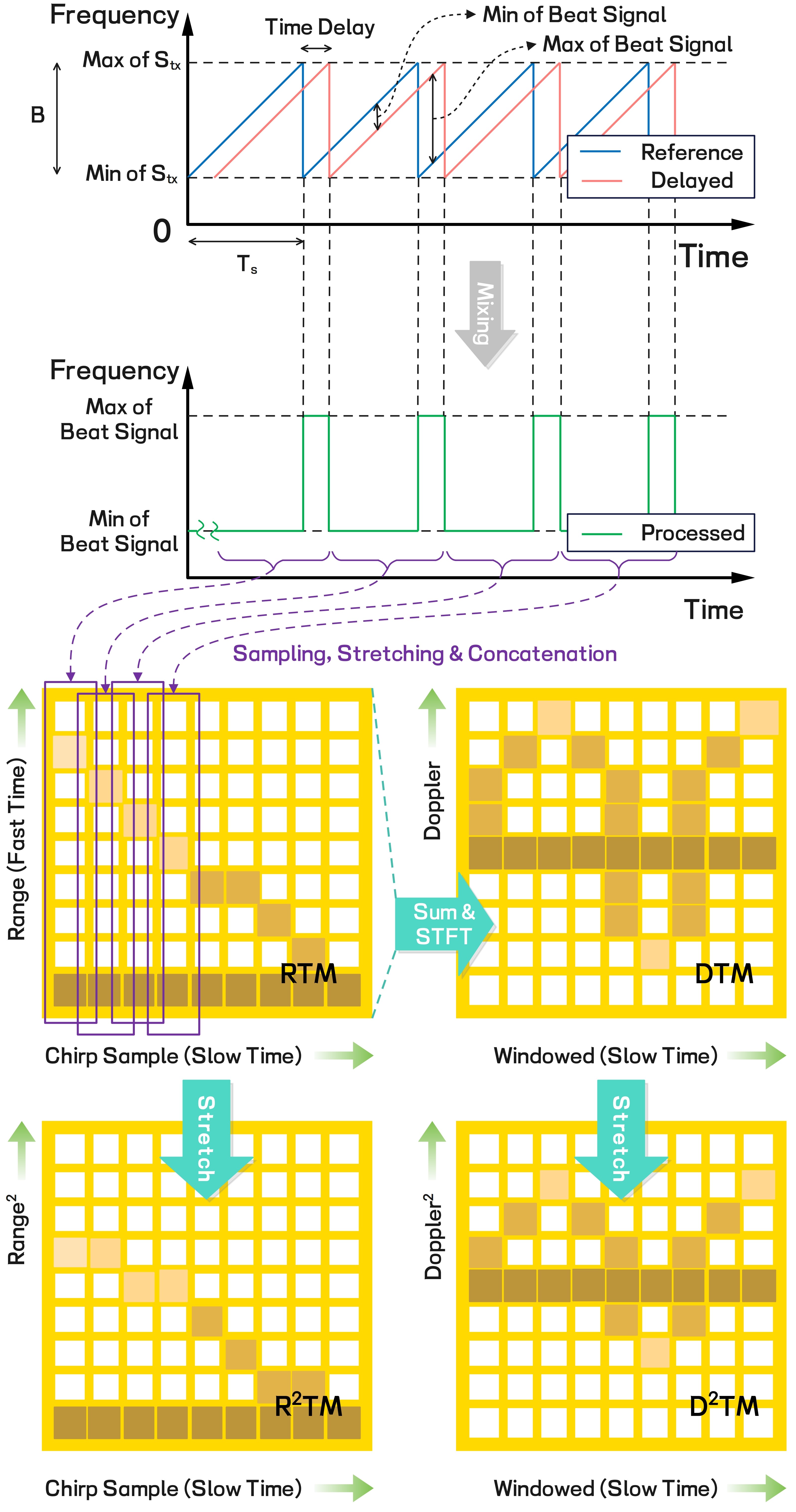}
    \caption{Principle of LFMCW signal transmitting, receiving, sampling, and time-frequency analysis.}
    \label{LFMCW}
    \vspace{-0.2cm}
\end{figure}\par
\begin{figure}[!ht]
\centering
\subfigure[]{\includegraphics[width=0.24\textwidth]{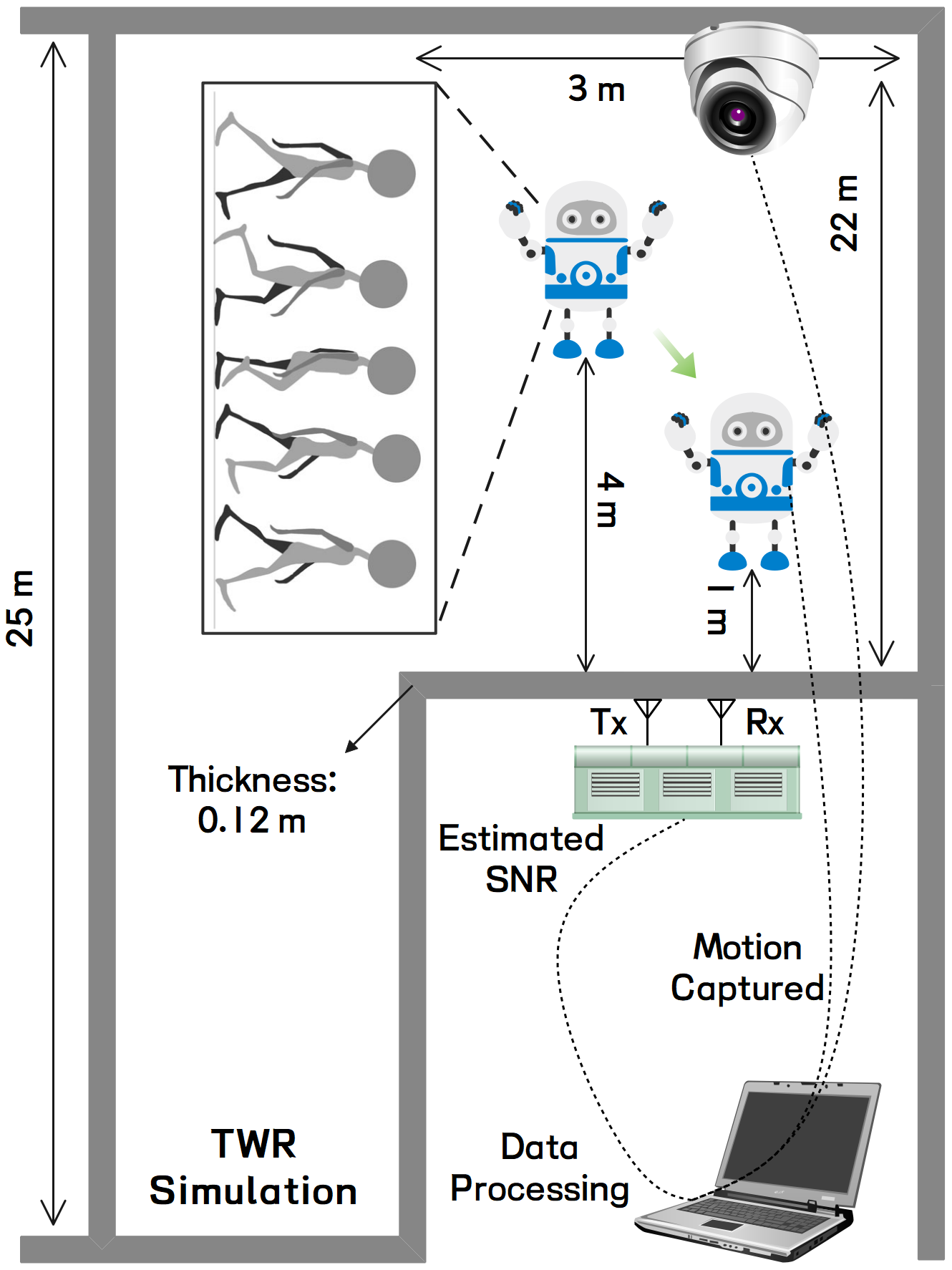}}
\subfigure[]{\includegraphics[width=0.24\textwidth]{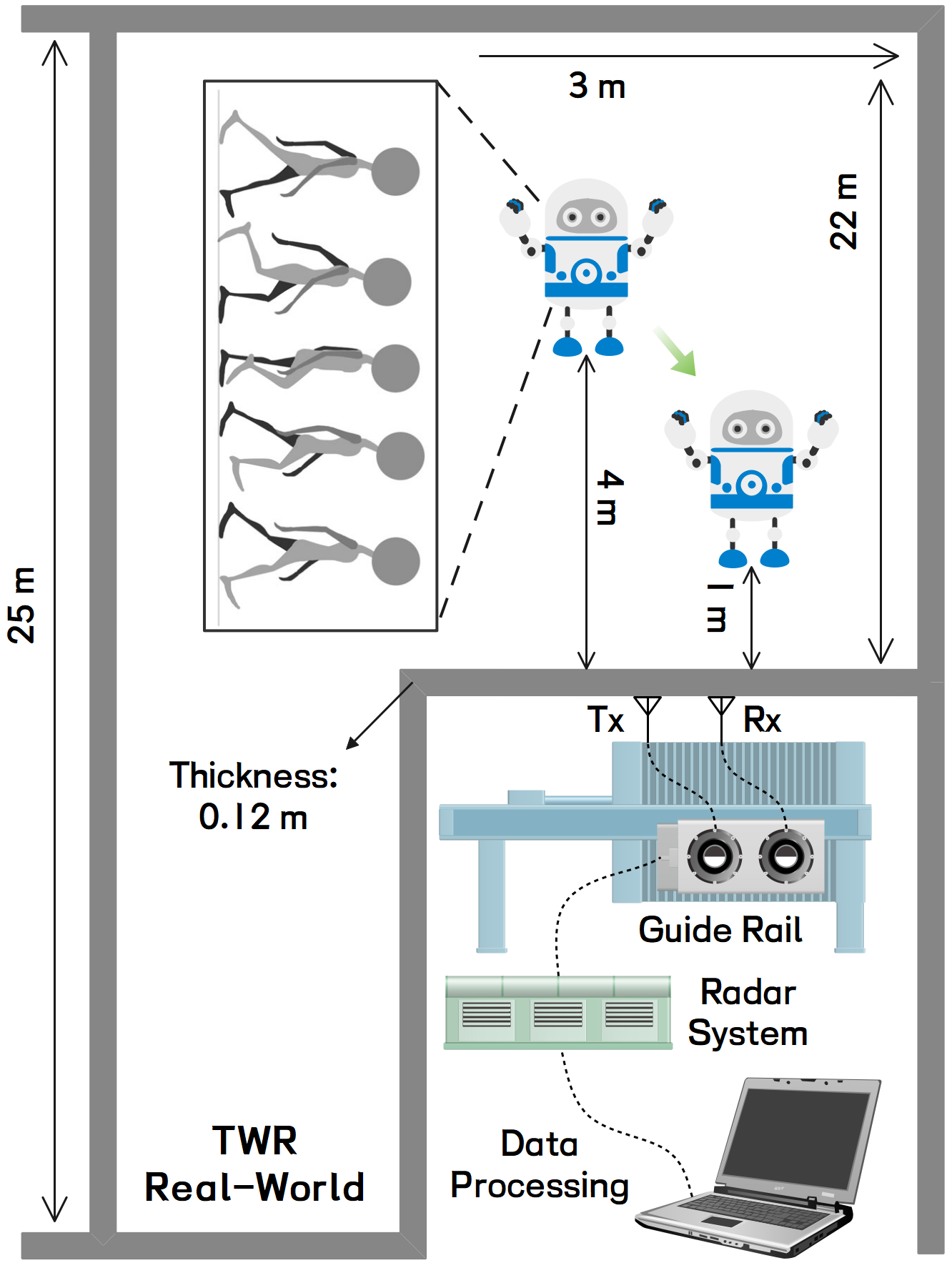}}
\caption{Schematic diagrams of the experimental scenarios: (a) through-the-wall, simulation, and (b) through-the-wall, real-world measurement. The motion captured data in (a) comes from open source works, while (b) is from the prototype TWR system we built ourselves.}
\label{Scene}
\vspace{-0.2cm}
\end{figure}\par
\subsection{Minimum Number of Corners in Range and T-F Profiles}
As shown in Fig. \ref{LFMCW}, to effectively display the micro-Doppler signature, the time-domain radar echoes are concatenated along the slow-time index $m$ to form a two-dimensional data matrix. After performing clutter and noise suppression on the matrix, the RTM is obtained. The clutter suppression is achieved using MTI filter and the noise suppression is achieved using EMD algorithm \cite{TWR-FMSN}.B y summing the data of all range cells in the RTM and utilizing STFT, a two-dimensional T-F image is obtained referred to as DTM \cite{VCChen}. The EMD algorithm is again used for clutter and noise residual suppression on the DTM \cite{TWR-FMSN}. From the analysis of the human motion model, it is found that the selection of points for reconstructing the complete information of the curves is related to the square of the range and velocity. Therefore, the RTM and DTM are stretched to square coordinates in vertical axes. The necessity lies in the fact that without this operation, the order of the resulting kinematic state curves for each limb node would change, and thus the minimum number of reconstruction corner points required would also change. The operation is achieved by interpolation, which finally outputs $\mathbf{R^2TM},\mathbf{D^2TM}$. Detailed flow for generating $\mathbf{R^2TM}$ and $\mathbf{D^2TM}$ is shown in Algorithm \ref{Squared in Vertical Axis}.\par
\begin{figure}[!ht]
    \centering
    \includegraphics[width=0.48\textwidth]{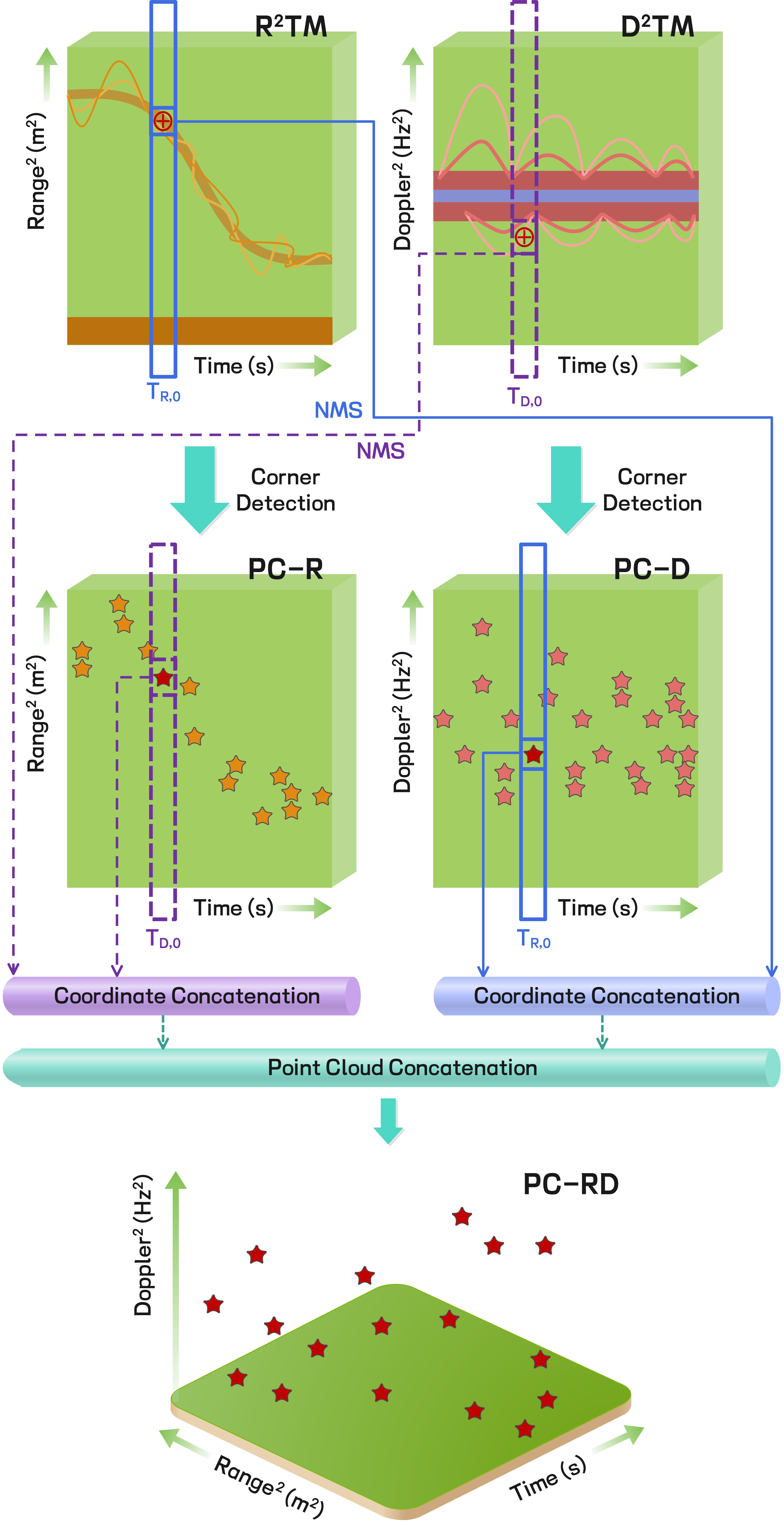}
    \caption{Schematic diagram of the generation process of $\mathbf{PC-RD}$.}
    \label{PC-RD Fusion}
    \vspace{-0.2cm}
\end{figure}\par
The horizontal and vertical coordinates corresponding to $\mathbf{R^2TM}$ and $\mathbf{D^2TM}$ consist of the trajectory equation obtained in human motion modeling. Corners in blob form usually appear at locations on the images where the curvature of the trajectory is the largest.\par
\begin{table*}
\begin{center}
\caption{Details of the Experiments$^{*}$.}
\label{Experimental Details}
\vspace{-0.4cm}
\resizebox{\textwidth}{!}{%
\begin{tabular}{ccc}
\hline\hline
\multicolumn{2}{c}{\textbf{Name of Validation Experiments}}             & \textbf{Interpretations}    \\ 
\hline
\multirow{3}{*}{Modeling Theory}   & Visualization &  Displaying radar images of the true label, simulated, and measured data with their corner representations \\
& Similarity & Calculating the similarity between the radar images, the corner representations, and the true values \\
& Robustness &  Testing the similarity between corner representations and their truth values under different noise levels \\
\multirow{3}{*}{Corner Representation Theory}   & Feature Embedding &  Comparing the feature separation level before and after corner representation \\
& Generalization Capability & Using frontier classifiers to compare the generalization ability before and after corner representation \\
& Scenario Adaptation &  Verifying the adaptation of the proposed method to inhomogeneous wall scenarios \\
\hline\hline
\end{tabular}%
}
\end{center}
\footnotesize $^{*}$ All these validations are analyzed in detail in subsections B and C of the experimental section.\\
\vspace{-0.2cm}
\end{table*}\par
\begin{table}[htbp]
\begin{center}
\caption{Uniform Parameters of Radar System$^{*}$.}\label{System Parameters}
\resizebox{0.48\textwidth}{!}{
\begin{tabular}{cc}
\hline\hline
\textbf{Parameters}             & \textbf{Value}     \\ 
\hline
Antenna Transceiver Spacing     & (SISO) $0.15 \mathrm{~m}$   \\
Work Center Frequency          & $1.5 \mathrm{~GHz}$   \\
Band Width                     & $2 \mathrm{~GHz}$  \\
Sampling Points$^{1}$ & $1024$                \\
Sampling Period & $4 \mathrm{~s}$                 \\
Wall Thickness & $0.12 \mathrm{~m}$  \\
Human Motion Range from Radar & $1 \sim 4 \mathrm{~m}$     \\
SNR of Raw Data$^{2}$ & $-19.85 \sim -12.46 \mathrm{~dB}$ \\
SNR of Processed Data$^{2}$ & $\approx0 \mathrm{~dB}$ \\
Antenna Height to Ground & $1.5 \mathrm{~m}$      \\ 
Number of Activities$^{3}$ & $12$ \\
\hline\hline
\end{tabular}
}
\end{center}
\footnotesize $^{*}$ For reasons of rigor, we unify the simulated and measured parameters.\\
\footnotesize $^{1}$ Both on fast-time and slow-time. Therefore the final generated time-domain echo matrix should be in square size.\\
\footnotesize $^{2}$ The SNR mentioned in the table are obtained by manually selecting the target's region of the image and calculating the image pixel energy \cite{SNR}.\\
\footnotesize $^{3}$ The categories and the amount of data might not correspond in some datasets. Thus, we reselect and reorganize the labels from the original and introduce overlap window slicing to ensure consistency of the experiments.\\
\vspace{-0.4cm}
\end{table}\par
Based on the results of the human motion model, the squared one-way propagation distance curves and squared velocity curves are summarized and rewritten for each scattering center on $\mathbf{R^2TM}$ and $\mathbf{D^2TM}$ in TABLE \ref{Minimum Corner Points}. The minimum number of corner points required to reconstruct the complete information for each curve is calculated separately. For activities where the body's position changes, the minimum required number of corner points on $\mathbf{R^2TM},~\mathbf{D^2TM}$ is $30$ and $22$, respectively. For activities where the body's position does not change, the minimum required number of corner points on both $\mathbf{R^2TM},~\mathbf{D^2TM}$ is $30$. Consider the tolerance, and for ease of algorithm design, the number of corners is selected as $30$.\par

\section{Experimental Verification}
In this section, the parameters of radar systems used for TWR HAR in both simulated and measured scenarios are first given, and the method for achieving feature extraction is explained. Then, as shown in TABLE \ref{Experimental Details}, the correctness and effectiveness of the modeling theory and the corner representation theory are verified in sequence. Finally, a discussion of the directions that remain to be optimized for proposed theory is given.\par

\subsection{Configuration of Radar System}
As shown in TABLE \ref{System Parameters}, parameters for the developed TWR systems are presented. Both simulation and real-world experiment data sets are collected for validation. In the following, the terms “Simulated” and “Measured” are used to refer to these two data sets, respectively. The radar is placed outdoors for indoor human detection. The height from radar antennas to the ground is $1.5~m$. The center frequency is of $1.5\mathrm{~GHz}$ with bandwidths of $2\mathrm{~GHz}$. The total number of fast time samples is $1024$. the number of slow time samples is $256$ per second, and the sampling time window is $4$ seconds, thus the resulting one frame of radar data is a square matrix. The thickness of both simulated and measured walls is $0.12~m$, where an isotropic rectangle with a relative dielectric constant of $6$ is used in simulated scenario instead. The range of indoor human activities is from $1\sim 4~m$. There are $12$ human activities in total, including: $S1$, Empty; $S2$, Punching; $S3$, Kicking; $S4$, Grabbing; $S5$, Sitting Down; $S6$, Standing Up; $S7$, Rotating; $S8$, Walking; $S9$, Sitting to Walking; $S10$, Walking to Sitting; $S11$, Falling to Walking; $S12$, Walking to Falling. All the rest parameters are kept consistent for the two radar systems.\par
The two data collection scenarios are schematically shown in Fig. \ref{Scene}. For simulated scenario, the data set is referenced from related open source work from UCL \cite{UCL}. The team utilizes the motion capture method to get the trajectory curves of the nodes of the human body during various activities. Based on the team’s data, the locations of each limb node at specific moments are obtained. The limb nodes are populated with isotropic homogeneous ellipsoids to generate radar echoes under through-the-wall scene using low-frequency UWB transceiver. To standardize the activity categories and sampling time window lengths, the datas generated from the public dataset are re-selected and re-sliced \cite{CI4R}.\par
For measured scenario, the data set is obtained by the experimental prototype of the TWR built in \cite{TWR-FMSN} with the parameters described above. The interpretability of the data is improved by pre-processing operations such as standards-based channel calibration. Similar to the operation of the simulated data set, to standardize the activity categories and sampling time window lengths, the datas collected are re-selected and re-sliced.\par
\begin{figure*}[!ht]
    \centering
    \includegraphics[width=\textwidth]{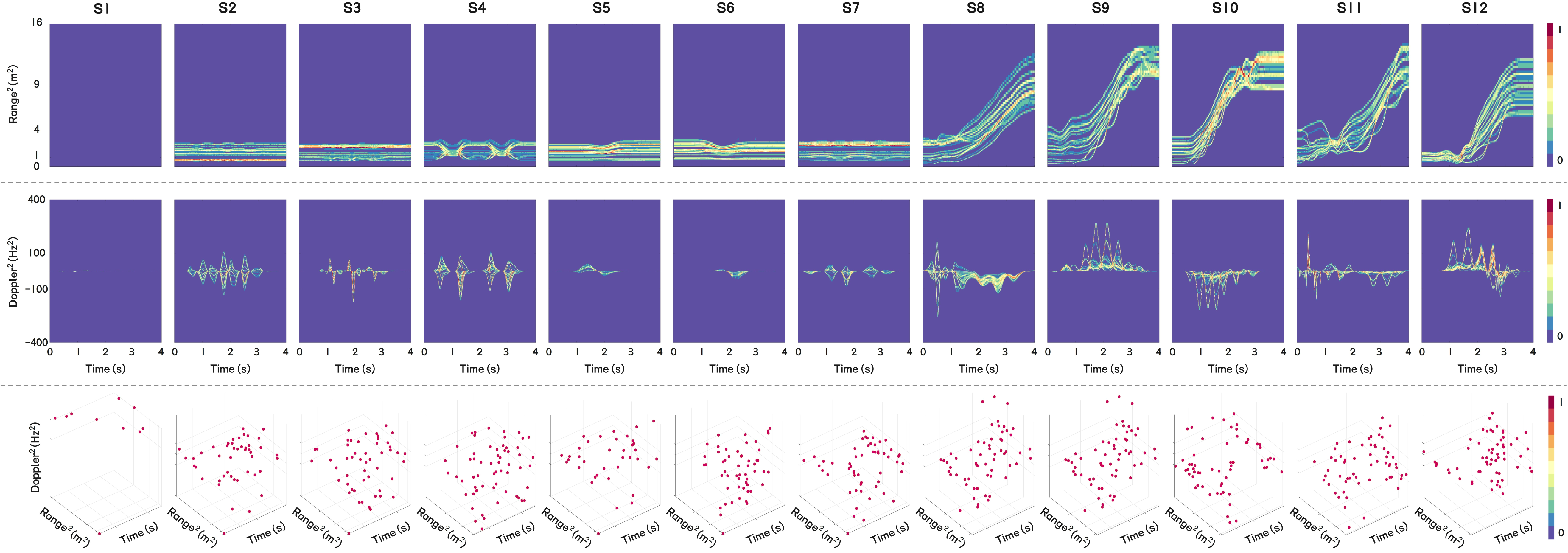}
    \caption{Visualization of groundtruth under $12$ human activities: The first row shows the $\mathbf{R^2TM}$ images, the second row shows the $\mathbf{D^2TM}$ images, and the third row shows the obtained micro-Doppler corner representation $\mathbf{PC-RD}$. Activity labels include: $S1$, Empty; $S2$, Punching; $S3$, Kicking; $S4$, Grabbing; $S5$, Sitting Down; $S6$, Standing Up; $S7$, Rotating; $S8$, Walking; $S9$, Sitting to Walking; $S10$, Walking to Sitting; $S11$, Falling to Walking; $S12$, Walking to Falling.}
    \label{GroundTruth Visualization}
    \vspace{-0.0cm}
\end{figure*}\par
\begin{figure*}[!ht]
    \centering
    \includegraphics[width=\textwidth]{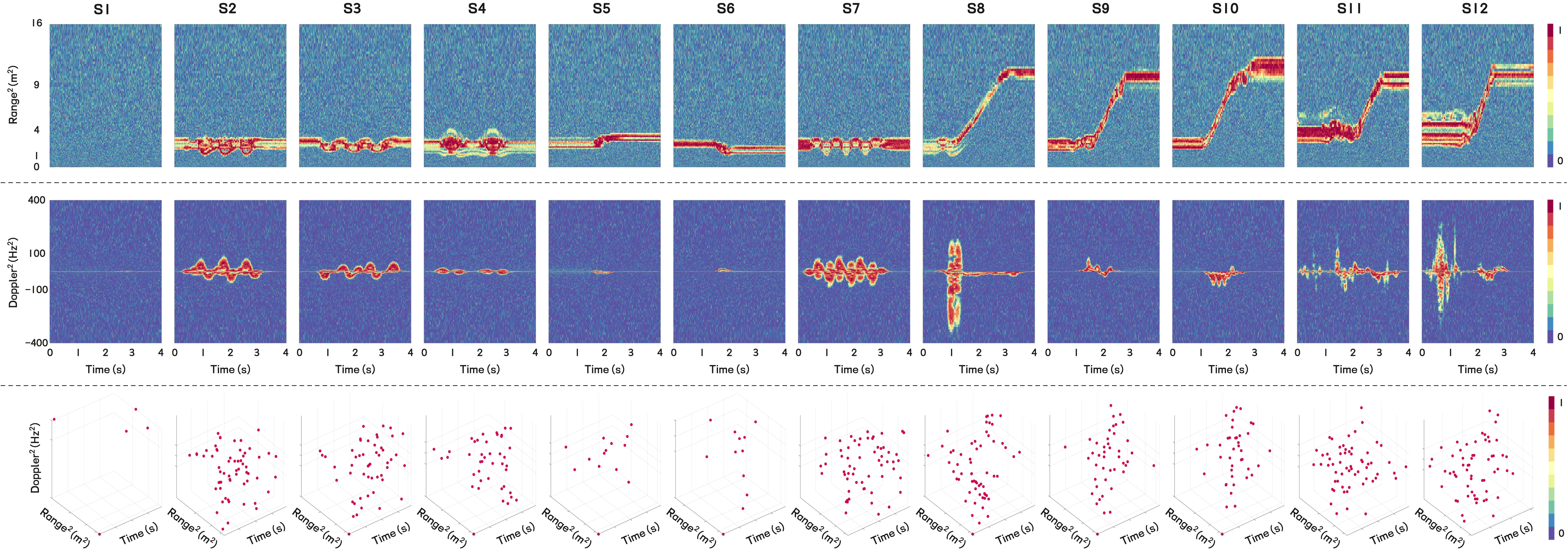}
    \caption{Visualization of simulated datas under $12$ human activities: The first row shows the $\mathbf{R^2TM}$ images, the second row shows the $\mathbf{D^2TM}$ images, and the third row shows the obtained micro-Doppler corner representation $\mathbf{PC-RD}$. $S1-S12$ are consistent with the definition in Fig. \ref{GroundTruth Visualization}.}
    \label{Simulated Visualization}
    \vspace{-0.0cm}
\end{figure*}\par
\begin{figure*}[!ht]
    \centering
    \includegraphics[width=\textwidth]{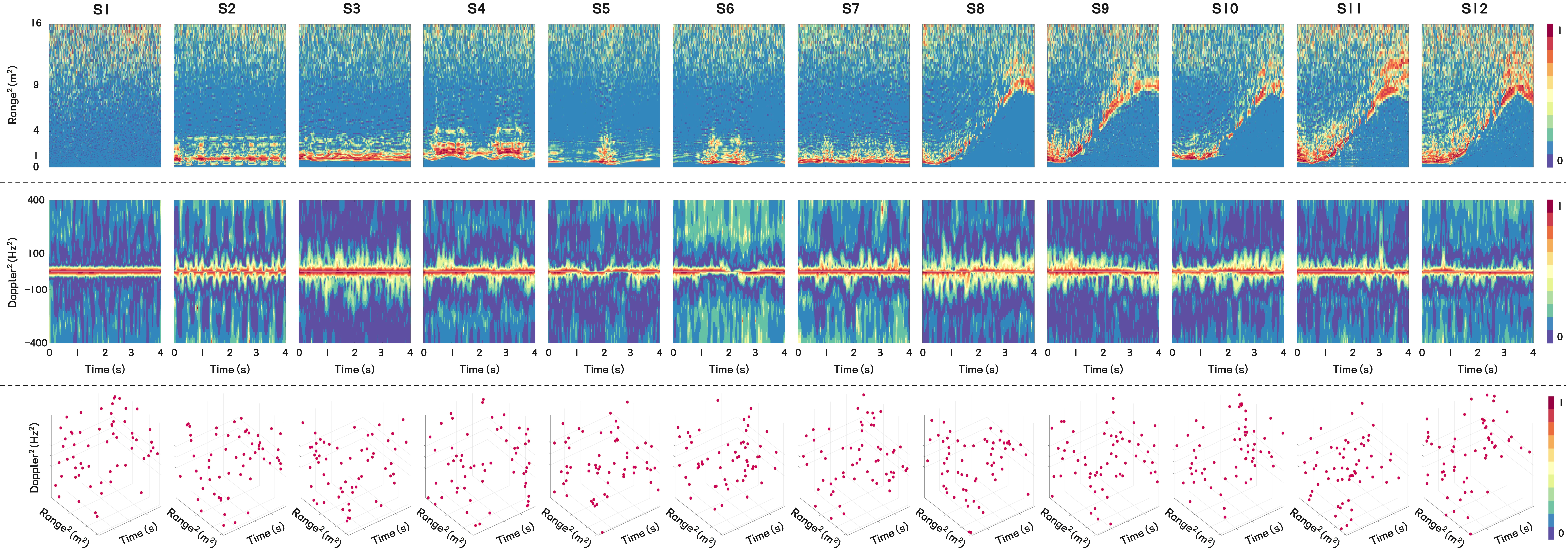}
    \caption{Visualization of measured datas under $12$ human activities: The first row shows the $\mathbf{R^2TM}$ images, the second row shows the $\mathbf{D^2TM}$ images, and the third row shows the obtained micro-Doppler corner representation $\mathbf{PC-RD}$. $S1-S12$ are consistent with the definition in Fig. \ref{GroundTruth Visualization}.}
    \label{Measured Visualization}
    \vspace{-0.0cm}
\end{figure*}\par
\begin{figure*}[!ht]
    \centering
    \includegraphics[width=\textwidth]{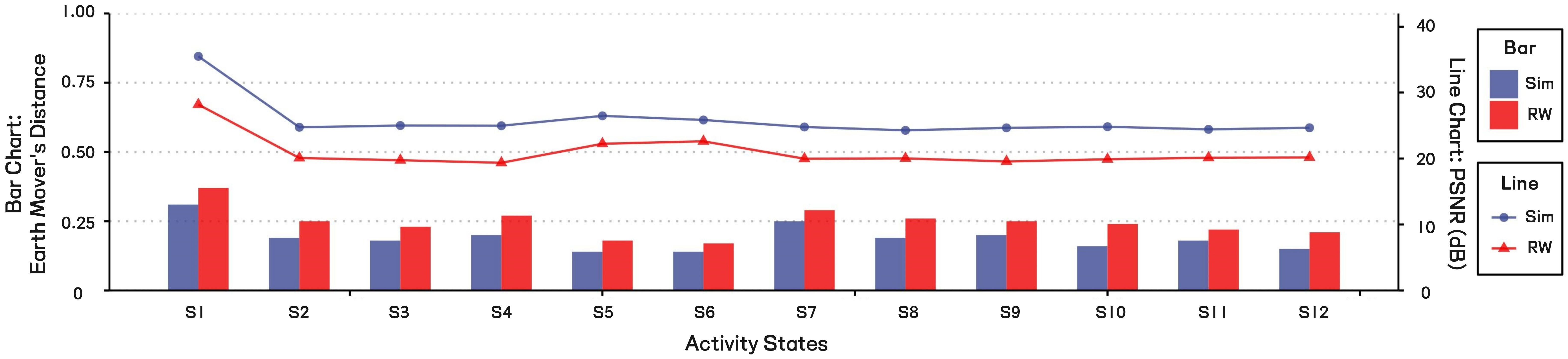}
    \caption{PSNR of the two types of $\mathbf{R^2TM}$ versus groundtruth, and the earth mover's distance of the corner representation versus the groundtruth point cloud. “Sim” is the abbreviation of “Simulated”. “RW” is the abbreviation of “Real-World Measured”. The remaining definitions are consistent with Fig. \ref{GroundTruth Visualization}.}
    \label{RTM EMD PSNR}
    \vspace{0.2cm}
\end{figure*}\par
\begin{figure*}[!ht]
    \centering
    \includegraphics[width=\textwidth]{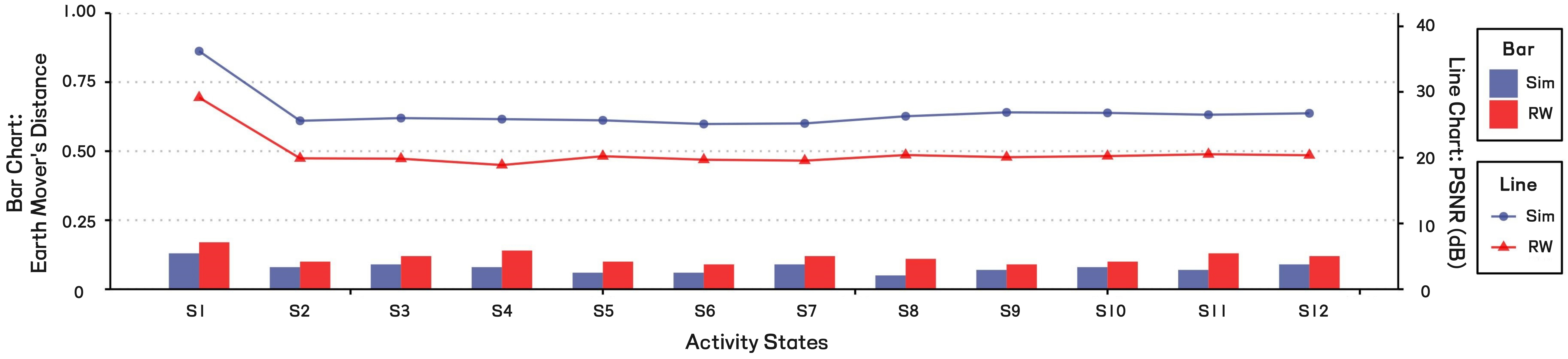}
    \caption{PSNR of the two types of $\mathbf{D^2TM}$ versus groundtruth, and the earth mover's distance of the corner representation versus the groundtruth point cloud. “Sim” is the abbreviation of “Simulated”. “RW” is the abbreviation of “Real-World Measured”. The remaining definitions are consistent with Fig. \ref{GroundTruth Visualization}.}
    \label{DTM EMD PSNR}
    \vspace{0.2cm}
\end{figure*}\par
\begin{figure*}[!ht]
    \centering
    \includegraphics[width=\textwidth]{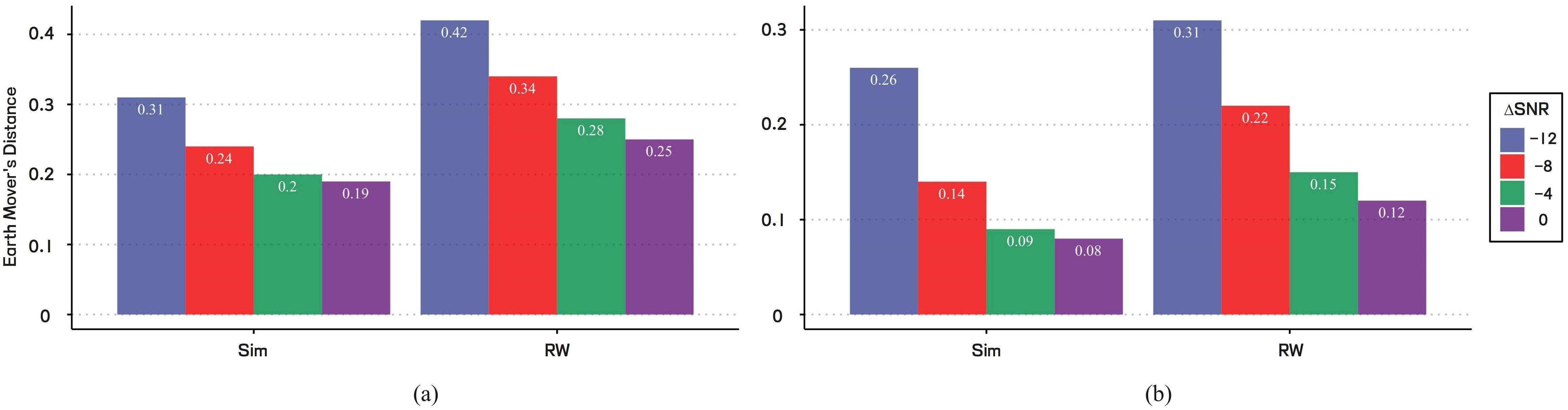}
    \caption{Robustness test of corner representations at different noise levels, where (a) is on $\mathbf{R^2TM}$ and (b) is on $\mathbf{D^2TM}$. $\Delta \mathrm{SNR}$ is the value of SNR decreasing in $\mathrm{dB}$ unit when adding different power of Gaussian noise manually. “Sim” is the abbreviation of “Simulated”. “RW” is the abbreviation of “Real-World Measured”. The remaining definitions are consistent with Fig. \ref{GroundTruth Visualization}.}
    \label{Robustness Plots}
    \vspace{-0.0cm}
\end{figure*}\par
In the following simulated and measured experiments, the SOGGDD filtering is used uniformly for blob-sensitive micro-Doppler corner detection on both $\mathbf{R^2TM}$ and $\mathbf{D^2TM}$ \cite{SOGGDD}. Define the point cloud data matrices obtained from corner detection using SOGGDD filtering on $\mathbf{R^2TM}$ and $\mathbf{D^2TM}$ as $\mathbf{PC-R}$ and $\mathbf{PC-D}$, respectively. Then, $\mathbf{PC-R}$ and $\mathbf{PC-D}$ are fused together to obtain a three-dimensional point cloud, which is used as the final extracted micro-Doppler signature. As shown in Fig. \ref{PC-RD Fusion}, take a corner point on one of the slow time of the $\mathbf{PC-R}$ and find the point of maximum amplitude on $\mathbf{D^2TM}$ for the data corresponding to the same slow time. This frequency coordinate is concatenated to the corner point on the $\mathbf{PC-R}$ to obtain a 3D coordinate. Similarly, take a corner point on one of the slow time of the $\mathbf{PC-D}$ and find the point of maximum amplitude on $\mathbf{R^2TM}$ for the data corresponding to the same slow time. This frequency coordinate is concatenated to the corner point on the $\mathbf{PC-D}$ to obtain a 3D coordinate. Iterating through both $\mathbf{PC-R}$ and $\mathbf{PC-D}$ and concatenating all the 3D coordinates into a matrix, which obtains the desired micro-Doppler corner point cloud representation $\mathbf{PC-RD}$ with the dimension of $60 \times 3$.\par
\begin{figure*}[!ht]
    \centering
    \includegraphics[width=\textwidth]{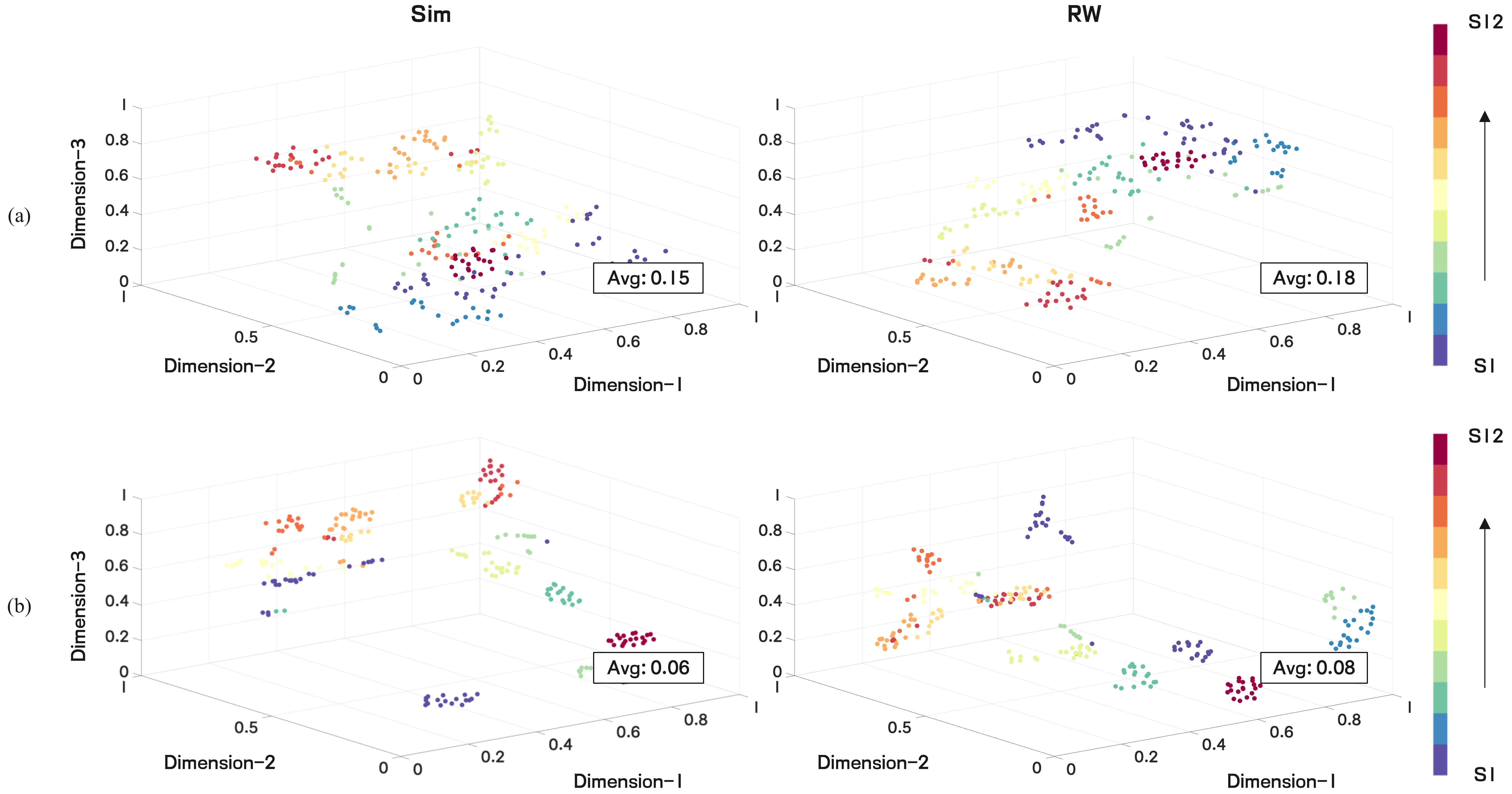}
    \caption{Results of T-SNE analysis on $\mathbf{R^2TM}$, where (a) is before corner representation, (b) is after corner representation. “Avg” is the abbreviation of “average distance from the center of clustering”. “Sim” is the abbreviation of “Simulated”. “RW” is the abbreviation of “Real-World Measured”. }
    \label{TSNE RTM}
    \vspace{0.2cm}
\end{figure*}\par
\begin{figure*}[!ht]
    \centering
    \includegraphics[width=\textwidth]{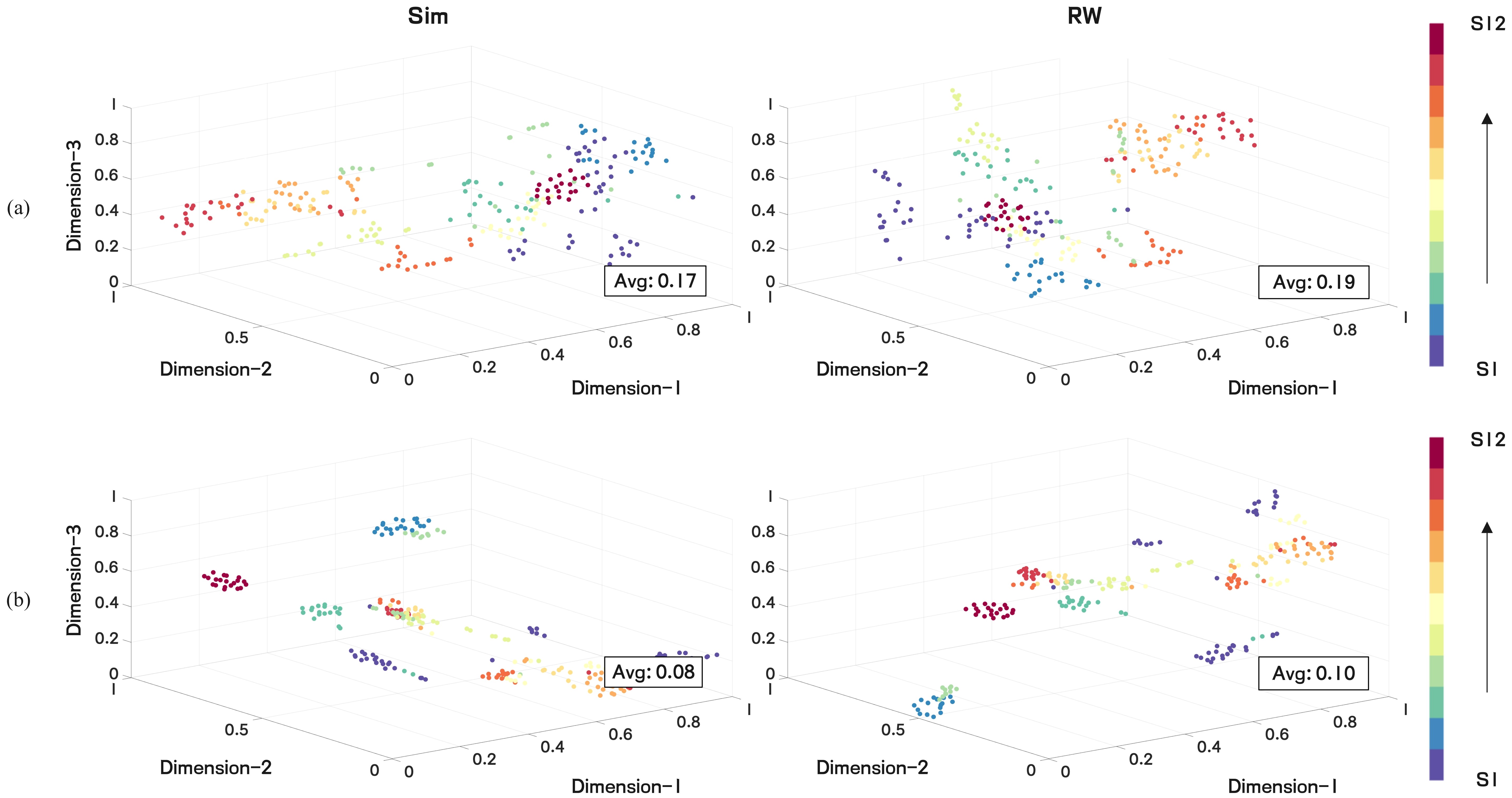}
    \caption{Results of T-SNE analysis on $\mathbf{D^2TM}$, where (a) is before corner representation, (b) is after corner representation. “Avg” is the abbreviation of “average distance from the center of clustering”. “Sim” is the abbreviation of “Simulated”. “RW” is the abbreviation of “Real-World Measured”. }
    \label{TSNE DTM}
    \vspace{-0.0cm}
\end{figure*}\par

\subsection{\textbf{Validation of Modeling Theory:} Visualization, Similarity, and Robustness}
The visualizations of our proposed micro-Doppler corner representation on two different types of TWR data are shown in Fig. \ref{Simulated Visualization} and Fig. \ref{Measured Visualization}, while the groundtruth maps generated using the reference trajectories of human motion obtained from the captured datas are shown in Fig. \ref{GroundTruth Visualization}. Fig. \ref{Simulated Visualization} gives the results for simulated scenario, and Fig. \ref{Measured Visualization} gives the results for measured scenario. Comparing Fig. \ref{Simulated Visualization} and Fig. \ref{Measured Visualization} with Fig. \ref{GroundTruth Visualization} can be a viable option for evaluating the effectiveness of the proposed model and feature representation method.\par
As shown in Fig. \ref{Simulated Visualization}, regardless of the type of activity, human motion simultaneously induces Doppler and micro-Doppler effects on both the generated $\mathbf{R^2TM}$ and $\mathbf{D^2TM}$. In Fig. \ref{Measured Visualization}, the Doppler and micro-Doppler signature tends to be much weaker, but still visible from the image results. The Doppler signature on the $\mathbf{R^2TM}$ is the overall trend of displacement of the nodes of the human body. The micro-Doppler signature, on the other hand, is the presence of micro-motion of the nodes of the body with respect to each other, which in turn results in the appearance of multiple peaks on or within the curve envelope. On the $\mathbf{D^2TM}$, the Doppler signature is the curves that are not axisymmetric or centrosymmetric. In contrast, the micro-Doppler signature is the appearance of peaks that are relatively weak in energy but have much higher frequencies. All these image features are used for corner representation.\par
Compared the $\mathbf{R^2TM}$ with groundtruth in both simulated and measured scenarios, it can be seen that the macroscopic motion trends of human limb nodes in the resulting images are basically the same. Compared $\mathbf{D^2TM}$ with groundtruth, it can be clearly seen that the relative motion trends of human limb nodes in the resulting images are basically the same.\par
The Doppler and micro-Doppler signature on the images acquired in simulated scenario are more obvious. The images acquired in measured scenario are even lower in SNR and resolution, which means that both Doppler and micro-Doppler signature are blurred. However, after corner extraction, both Doppler and micro-Doppler signature are well represented.\par
\begin{table}[!t]
\begin{center}
\caption{Comparison of the impact on the generalization capability of back-end recognition models on new testers' data before and after corner representation$^{*}$.}\label{Comparison Acc}
\vspace{-0.0cm}
\resizebox{0.48\textwidth}{!}{
\begin{tabular}{ccccccccc}
\hline\hline
\multirow{2}{*}{($\%$)}  & \multicolumn{4}{c}{\textbf{Sim}} & \multicolumn{4}{c}{\textbf{RW}}\\
\cline{2-9}& \textbf{Tr} & \textbf{Va} & \textbf{Te1} & \textbf{Te2}& \textbf{Tr} & \textbf{Va} & \textbf{Te1} & \textbf{Te2}\\
\hline
\multicolumn{9}{c}{\textbf{Methods Before Corner Representation}}\\
\hline
M1$^{1}$  &$95.41$	&$94.00$	&$80.50$	&$75.75$	&$91.66$	&$91.25$ 
    &$81.75$	&$71.25$
\\            
M2$^{2}$  &$95.44$	&$92.00$	&$79.50$	&$71.25$	&$90.28$	&$85.75$ 
    &$77.00$	&$70.50$
\\  
M3$^{3}$  &$90.84$	&$85.63$	&$74.00$	&$65.50$	&$88.47$	&$79.38$ 
    &$74.75$	&$60.25$
\\  
M4$^{4}$  &$98.97$	&$94.13$	&$80.00$	&$72.25$	&$97.72$	&$92.88$ 
    &$80.50$	&$66.25$
\\  
\hline 
\multicolumn{9}{c}{\textbf{\ding{79} Methods After Corner Representation \ding{79}}}\\
\hline
M1$^{1}$     &$93.22$	&$90.63$	&$88.00$	&$82.25$	&$92.84$	&$90.13$	&$86.50$	&$80.25$                  \\
M2$^{2}$   &$96.78$	&$93.38$	&$90.50$	&$84.75$	&$93.34$	&$87.63$	&$83.50$	&$80.00$                  \\
M3$^{3}$   &$93.91 $	&$90.75$	&$86.50$	&$81.00$	&$90.44$	&$84.63$	&$79.75$	&$71.25$                  \\
M4$^{4}$   &$97.28$	&$94.13$	&$89.50$	&$81.75$	&$96.06$	&$90.00$	&$85.75$	&$78.75$                  \\
\hline\hline
\end{tabular}
}
\end{center}
\footnotesize $^{*}$ "Tr", "Va", "Te1", and "Te2" are the abbreviations of "training accuracy", "validation accuracy", "accuracy on test set with $1.7~m$ human height", and "accuracy on test set with $1.6~m$ human height", respectively. The remaining definitions are consistent with Fig. \ref{GroundTruth Visualization}. In addition, to ensure the consistency and persuasiveness of the experiments, the image data used before the corner point representation are RTM and DTM, and the image data used after the corner point representation are $\mathbf{R^2TM}$ and $\mathbf{D^2TM}$.\\
\footnotesize $^{1}$ M1, SIMFNet \cite{SIMFNet}: The network of the authors' proposed method is equipped with three parallel input branches. However, considering the consistency of the comparison experiments, we remove the link of cadence velocity diagram input, and only use RTM / DTM braches for training.\\
\footnotesize $^{2}$ M2, DRDSP-PointNet \cite{DRDSP-PointNet}: The method originally has the process of feature dimension reduction, here we refer to the method proposed by the authors that also does the data sparse compression based on OMP. However, the data after corner representation does not perform the sparse compression process.\\
\footnotesize $^{3}$ M3, TWR-MDFF \cite{TWR-MDFF}: The data format in the feature extraction stage and the FastPCA fusion method both refer to the original paper. The concatenated dimension of “Data Index” is selected to have a dimension of $1024~(256+256+512)$.\\
\footnotesize $^{4}$ M4, TWR-RV-ConvGRU \cite{TWR-RV-ConvGRU}: For the rigor of comparison, the number of nodes in the ConvGRU layer used to extract Range-Time-Doppler data in the proposed method is still set to $60$, which is consistent with M1.\\
\vspace{-0.4cm}
\end{table}\par
In order to quantitatively assess that the corner representation do reflect the Doppler and micro-Doppler signature in both the $\mathbf{R^2TM}$ and $\mathbf{D^2TM}$, fully resolved earth mover's distance is introduced to measure the variability between two point clouds \cite{Earth Mover's Distance}. The earth mover's distance metric $\mathrm{E-M-Dist}$ of the $\mathbf{PC-R}$ and $\mathbf{PC-D}$ is:

\vspace{-0.4cm}
\begin{equation}
\text{E-M-Dist}(\mathbf{GT},\mathbf{Ta})=\frac{\sum_{i=1}^\zeta \sum_{j=1}^{\zeta^{\prime}} \mathrm{fp}_{ij}\mathrm{Dp}_{ij}}{\sum_{i=1}^\zeta \sum_{j=1}^{\zeta^{\prime}} \mathrm{fp}_{ij}},
\end{equation}
where $\zeta=\zeta^{\prime}=30$. $\mathbf{fp}=[\mathrm{fp}_{ij}]$, $\mathrm{fp}_{i j} \geq 0,\quad 1 \leq i \leq \zeta,~1 \leq j \leq {\zeta^{\prime}}$ is the corresponding weights, which can be obtained by solving:

\vspace{-0.2cm}
\begin{equation}
F_p=\arg\max_{F_p}\left(\sum_{i=1}^\zeta\sum_{j=1}^{\zeta^{\prime}}\mathrm{fp}_{ij}\mathrm{Dp}_{ij}\right),
\end{equation}
which satisfies the constraints below:

\vspace{-0.2cm}
\begin{equation}
\sum_{j=1}^{\zeta^{\prime}} \mathrm{fp}_{i j} \leq \mathrm{GT}_{\mathrm{po}_i}, \quad 1 \leq i \leq \zeta,
\end{equation}

\vspace{-0.2cm}
\begin{equation}
\sum_{i=1}^\zeta \mathrm{fp}_{i j} \leq \mathrm{Ta}_{{\mathrm{po}^{\prime}}_j}, \quad 1 \leq j \leq {\zeta^{\prime}},
\end{equation}

\vspace{-0.2cm}
\begin{equation}
\sum_{i=1}^\zeta \sum_{j=1}^{\zeta^{\prime}} \mathrm{fp}_{i j}=\min \left(\sum_{i=1}^\zeta \mathrm{GT}_{\mathrm{po}_i}, \sum_{j=1}^{\zeta^{\prime}} \mathrm{Ta}_{{\mathrm{po}^{\prime}}_j}\right), 
\end{equation}
where the groundtruth point cloud is:

\vspace{-0.3cm}
\begin{equation}
\mathbf{GT}=\{(\mathrm{po}_1,\mathrm{GT}_{\mathrm{po}_1}),\cdots,(\mathrm{po}_\zeta,\mathrm{GT}_{\mathrm{po}_\zeta})\},   
\end{equation}
and the target point cloud is:

\vspace{-0.4cm}
\begin{equation}
\mathbf{Ta}=\{({\mathrm{po}^{\prime}}_1,\mathrm{Ta}_{{\mathrm{po}^{\prime}}_1}),\cdots,({\mathrm{po}^{\prime}}_{\zeta^{\prime}},\mathrm{GT}_{{\mathrm{po}^{\prime}}_{\zeta^{\prime}}})\},    
\end{equation}
where $(\mathrm{po}_i,\mathrm{GT}_{\mathrm{po}_i}),~1 \leq i \leq \zeta$ and $({\mathrm{po}^{\prime}}_j,\mathrm{Ta}_{{\mathrm{po}^{\prime}}_j}),~1 \leq j \leq {\zeta^{\prime}}$ are normalized coordinates. The distance matrix is defined as $\mathbf{Dp}=[\mathrm{Dp}_{ij}]$, where $\mathrm{Dp}_{ij}$ denotes the Euclidean distance between point $(\mathrm{po}_i,\mathrm{GT}_{\mathrm{po}_i})$ and point $({\mathrm{po}^{\prime}}_j,\mathrm{Ta}_{{\mathrm{po}^{\prime}}_j})$. The unit of E-M-Dist is pixel $1$. In \cite{PCN}, if the earth mover's distance is less than $1$, it proves that the data is valid. If the earth mover's distance is less than $0.5$, it proves that the point cloud possesses similarity. If the earth mover's distance is less than $0.25$ for the vast majority of the data, it proves that the method has a good performance.\par
Fig. \ref{RTM EMD PSNR} and \ref{DTM EMD PSNR} give the earth mover's distance and PSNR results under different human activities. The calculation of PSNR on radar images can be found in paper \cite{TWR-MCAE}. The barchart denotes the earth mover's distance versus $12$ human activities in unit $1$, which locates on the left side of the vertical axis. The linechart denotes the PSNR versus $12$ human activities in unit $\mathrm{dB}$ \cite{PSNR}, which locates on the right side of the vertical axis.\par
From the results, for the group of empty scene ($S1$), both two types of $\mathbf{R^2TM}$ or $\mathbf{D^2TM}$ are relatively close to the groundtruth. These profiles do not exhibit any Doppler or micro-Doppler signature associated with human motion. Therefore, there is a relatively large disparity in the corner representation. For the other $11$ categories of activities, the earth mover's distance of the corner representation to the groundtruth is all less than $0.3$, which implies that the corner point cloud can represent the Doppler and micro-Doppler signature related to human motion \cite{PCN}. On $\mathbf{R^2TM}$, the order of PSNRs for two different data sets is: Simulated $>$ Measured. All the PSNRs are higher than $19 \mathrm{~dB}$. On $\mathbf{D^2TM}$, the order of PSNRs for two different data sets is also: Simulated $>$ Measured. All the PSNRs are higher than $18.5 \mathrm{~dB}$. From the above analysis, the extracted corner feature maps can reflect approximately human motion patterns.\par
Next, the noise robustness of the proposed micro-Doppler corner representation is verified by manually adding Gaussian noise of different powers to the $\mathbf{R^2TM}$ and $\mathbf{D^2TM}$. The SNRs of the images are decreased by $4\mathrm{~dB}$, $8\mathrm{~dB}$, and $12\mathrm{~dB}$. The results for earth mover's distance are shown in Fig. \ref{Robustness Plots}, which are calculated by arithmetically averaging over $12$ different activities. The results show that the distance is not significantly degraded. For relatively noisy environments that the reduction of SNR exceeds $10\mathrm{~dB}$, the distance is still below $0.5$, which is within an acceptable level \cite{PCN}.\par

\subsection{\textbf{Validation of Corner Representation Theory:} Feature Embedding, Generalization Capability, and Scenario Adaptation}
First, T-SNE is employed to analyze the feature separability of different activities in two different scenarios before and after the corner representation. T-SNE visualizes the data by mapping the high-dimensional profiles to the low-dimensional space while maintaining the local similarity relationship between data points \cite{T-SNE}. The results are shown in Fig. \ref{TSNE RTM} and \ref{TSNE DTM}. From the results, the inter-class feature separation of $\mathbf{R^2TM}$ and $\mathbf{D^2TM}$ obtained from simulated scenario is larger than measured scenario. After corner extraction and representation, the average distance of the intra-class data are all reduced compared to the original profiles while keeping the coordinate scale of the mapping space constant in $[0,1]$. Furthermore, compared with the results of the raw $\mathbf{R^2TM}$ and $\mathbf{D^2TM}$, the corner feature maps show larger inter-class distance and obvious clustering center. More importantly, the outstanding characteristics of corner representation makes potential benefits for high recognition accuracy and generalization capability.\par
Next, the generalization capability of the proposed micro-Doppler corner representation method is demonstrated by introducing some existed back-end recognition methods. The data sets are divided into training sets, validation sets and test sets. The training sets are used for model training, the validation sets are used for tuning the hyperparameters of the model and for model selection, and the test sets are used to evaluate the performance of the models. Note that, the testers employed in validation sets are the same as those in training sets, while the testers employed in test sets are different. Thus, the difference of the recognition accuracy between the test sets and validation sets reflect the model's generalization ability on different testers. The training and validation sets are generated based on two different testers with $1.8~m$ in height. The test sets are generated based on two different testers with $1.7~m$ and $1.6~m$ in height. The $\mathbf{R^2TM}$ and $\mathbf{D^2TM}$ for each scene are randomly sliced into $3200$ training samples, $800$ validation samples, and $400$ test samples, where activity state $\mathrm{S1}$ accounts for $17.5\%$ of the total data, and $\mathrm{S2}\sim \mathrm{S12}$ each accounts for $7.5\%$ of the total data. The amount of data for different tester identities is the same for all types of activities.\par
\begin{table}[!t]
\begin{center}
\caption{Comparison of the impact on the generalization capability of back-end recognition models under different wall scenarios$^{*}$.}\label{Comparison Wall}
\vspace{-0.0cm}
\resizebox{0.48\textwidth}{!}{
\begin{tabular}{ccccccccc}
\hline\hline
\multirow{2}{*}{($\%$)}  & \multicolumn{4}{c}{\textbf{Sim}} & \multicolumn{4}{c}{\textbf{Sim Inhomogeneous Wall$^{1}$}}\\
\cline{2-9}& \textbf{Tr} & \textbf{Va} & \textbf{Te1} & \textbf{Te2}& \textbf{Tr} & \textbf{Va} & \textbf{Te1} & \textbf{Te2}\\
\hline
\multicolumn{9}{c}{\textbf{Methods Before Corner Representation}}\\
\hline
M1  &$95.41$	&$94.00$	&$80.50$	&$75.75$	&$93.25$	&$90.88$ 
    &$86.50$	&$78.00$
\\            
M2  &$95.44$	&$92.00$	&$79.50$	&$71.25$	&$93.59$	&$86.38$ 
    &$83.25$	&$73.25$
\\  
M3  &$90.84$	&$85.63$	&$74.00$	&$65.50$	&$89.00$	&$88.50$ 
    &$75.50$	&$68.25$
\\  
M4  &$98.97$	&$94.13$	&$80.00$	&$72.25$	&$95.47$	&$93.63$ 
    &$85.00$	&$79.25$
\\  
\hline 
\multicolumn{9}{c}{\textbf{\ding{79} Methods After Corner Representation \ding{79}}}\\
\hline
M1     &$93.22$	&$90.63$	&$88.00$	&$82.25$	&$91.16$	&$90.63$	&$87.25$	&$81.00$                  \\
M2   &$96.78$	&$93.38$	&$90.50$	&$84.75$	&$94.97$	&$92.38$	&$89.25$	&$84.75$                  \\
M3   &$93.91 $	&$90.75$	&$86.50$	&$81.00$	&$92.38$	&$88.25$	&$83.50$	&$78.50$                  \\
M4   &$97.28$	&$94.13$	&$89.50$	&$81.75$	&$96.47$	&$93.75$	&$88.00$	&$80.50$                  \\
\hline\hline
\end{tabular}
}
\end{center}
\footnotesize $^{*}$ All abbreviations and parameter settings are consistent with TABLE \ref{Comparison Acc}.\\
\footnotesize $^{1}$ According to the experimental setup, the original dataset is constructed using a homogeneous wall. The dataset used for the comparison, on the other hand, considers an inhomogeneous wall setup consistent with the work \cite{Hao Ling 2}.\\
\vspace{-0.0cm}
\end{table}\par
\begin{figure}[!t]
    \centering
    \includegraphics[width=0.5\textwidth]{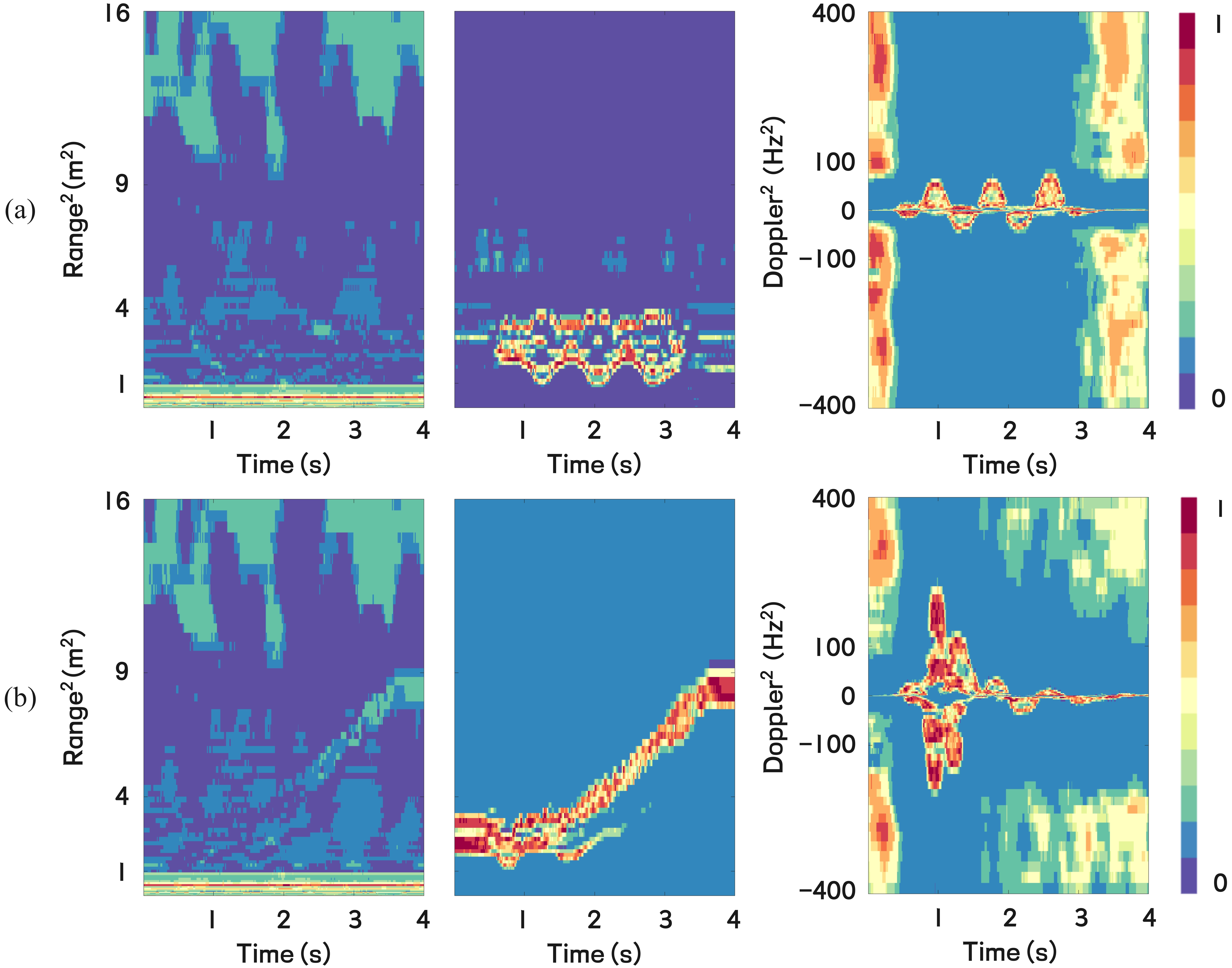}
    \caption{Simulated $\mathbf{R^2TM}$ and $\mathbf{D^2TM}$ under inhomogeneous wall \cite{Hao Ling 2, GPRMax}: (a) Punching activity, (b) Walking activity as an example. The first row presents the raw simulation in the form of  $\mathbf{R^2TM}$. The second row presents the processed $\mathbf{R^2TM}$. The third row presents the processed $\mathbf{D^2TM}$.}
    \label{Complex Wall Simulation}
    \vspace{-0.2cm}
\end{figure}\par
The selected recognition models include four frontier works: (1) M1: SIMFNet, (2) M2: DRDSP-PointNet, (3) M3: TWR-MDFF, and (4) M4: TWR-RV-ConvGRU. In all recognition models, the training optimizer keeps consistent with each other and Adam is used with the initial learning rate $0.00147$. The training process is continued until convergence and the model of the last epoch is used for inference.\par
The recognition accuracy of different methods is shown in TABLE \ref{Comparison Acc}. The training accuracy exceeds $90\%$, and validation accuracy exceeds $85\%$, which proves that these methods can achieve convergence in both simulated and measured scenarios, and the inference results are valid. The test accuracy of different recognition models using RTM, DTM as input drops heavily compared with validation accuracy. From the perspective of modeling, SIMFNet, TWR-MDFF, and TWR-RV-ConvGRU that do convolution, multi-scale conversion, multi-link fusion, or attention-based feature extraction on images have certain degree of generalization capability. DRDSP-PointNet that inherently has feature dimension reduction achieves relatively better generalization performance. By comparing the recognition accuracy under different testers, i.e. Te1 $>$ Te2, it can be concluded that the Doppler and micro-Doppler signature are heavily influenced by the testers' height. All the methods suffers from an accuracy degradation of more than $15\%$ when there is a large difference between height (Va versus Te2).\par
After corner representation, the performance of most of these methods on the training, validation, and test sets improves to a certain degree. The accuracy gaps between Va, Te1, and Te2 are reduced, which shows that corner representation can effectively focus on the key information of human motion micro-Doppler signature and improves generalization capability to testers with different height.\par
The performance of the generalization ability of the proposed method under different complex wall scenarios is also verified. In this case, both homogeneous and inhomogeneous walls and data generation methods refer to the work \cite{Hao Ling 2}. The signal model and the parameters of the TWR system used are kept consistent with the experimental setup described earlier in this paper. Examples of data visualization based on the inhomogeneous wall simulation is shown in Fig. \ref{Complex Wall Simulation} \cite{Wall Improvement 1, Wall Improvement 2, Wall Improvement 3, GPRMax}, and the comparisons of the accuracy of different existing recognition methods is shown in TABLE \ref{Comparison Wall}. From the results, it can be seen that the presence of a inhomogeneous wall brings amplitude distortion to the micro-Doppler signature of indoor human motion. The amplitude distortion introduced by inhomogeneous wall with respect to homogeneous wall is reduced to almost zero only after the clutter and noise suppression, and the normalization are performed. However, the curvature and Doppler frequency of the trace corresponding to each human limb nodes do not change. Thus the positions of the corner features analyzed from the kinematic model should not change. Comparing the validation and testing accuracy of the existing methods before and after the corner representation, it can be seen that the proposed method remains good generalization ability even in the presence of inhomogeneous wall.\par

\subsection{Discussions}
The effectiveness and generalization ability of the proposed method are proved both mathematically and experimentally. However, there are still some limitations and potentials, which are worth investing in further research, including:\par
\textbf{1. Limitations on Model Complexity:} Although the lowest feature dimension constraints and simplicity that can be computed through the proposed Boulic-sinusoidal pendulum kinematic model, it is still flawed for portraying the complex stochasticity of real human motion. Modeling studies that trade-off between simplicity and realism are still necessary.\par
\textbf{2. Potentials on the Method of Calculating the Optimal Corner Points}: The paper proposes a method using the summation of the order information of the nodal curves in the kinematic model. The method shows theoretical soundness and can be analyzed in depth for their respective practical value in further researches.\par

\section{Conclusion}
In this paper, we have proposed a corner-based micro-Doppler signature representation method to address the issue of accuracy loss for various testers. This paper has provided the joint Boulic-sinusoidal pendulum model to characterize indoor human activities as well as a refined signal model of TWR echoes. Additionally, the paper has analyzed the micro-Doppler corner feature for human motion on $\mathbf{R^2TM}$ and $\mathbf{D^2TM}$, and computed the minimum number of corner points needed to describe the Doppler and micro-Doppler information. Furthermore, we have given a feasible corner extraction method based on SOGGDD filter. Both simulated and measured experiments are conducted to verify the proposed theory. The findings have shown that the suggested micro-Doppler corners could accurately represent the motion characteristics of human limb nodes, which shows great benefits for high recognition accuracy and generalization capability. Our future work will focus on proposing a more accurate and intelligent micro-Doppler corner detection and activity recognition method.\par



\newpage
\begin{IEEEbiography}[{\includegraphics[width=1in,height=1.25in,clip,keepaspectratio]{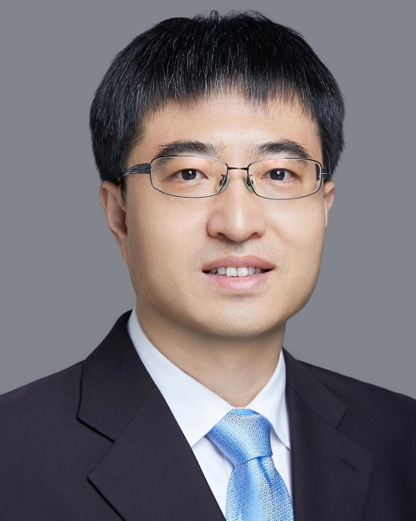}}]{Xiaopeng Yang}
Xiaopeng Yang (Senior Member, IEEE), professor, received B.E. and M.E. degrees from Xidian University, China, in 1999 and 2002, respectively, and Ph.D. degree from Tohoku University, Japan, in 2007. He was a Post-Doctoral Research Fellow with Tohoku University from 2007 to 2008 and a Research Associate with Syracuse University, USA, from 2008 to 2010. Since 2010, he has been working with the School of Information and Electronics, Beijing Institute of Technology, Beijing, China, where he is currently a full professor.\par
His current research interests include the phase array radar, ground-penetrating radar and through-the-wall radar.\par
\end{IEEEbiography}
\begin{IEEEbiography}[{\includegraphics[width=1in,height=1.25in,clip,keepaspectratio]{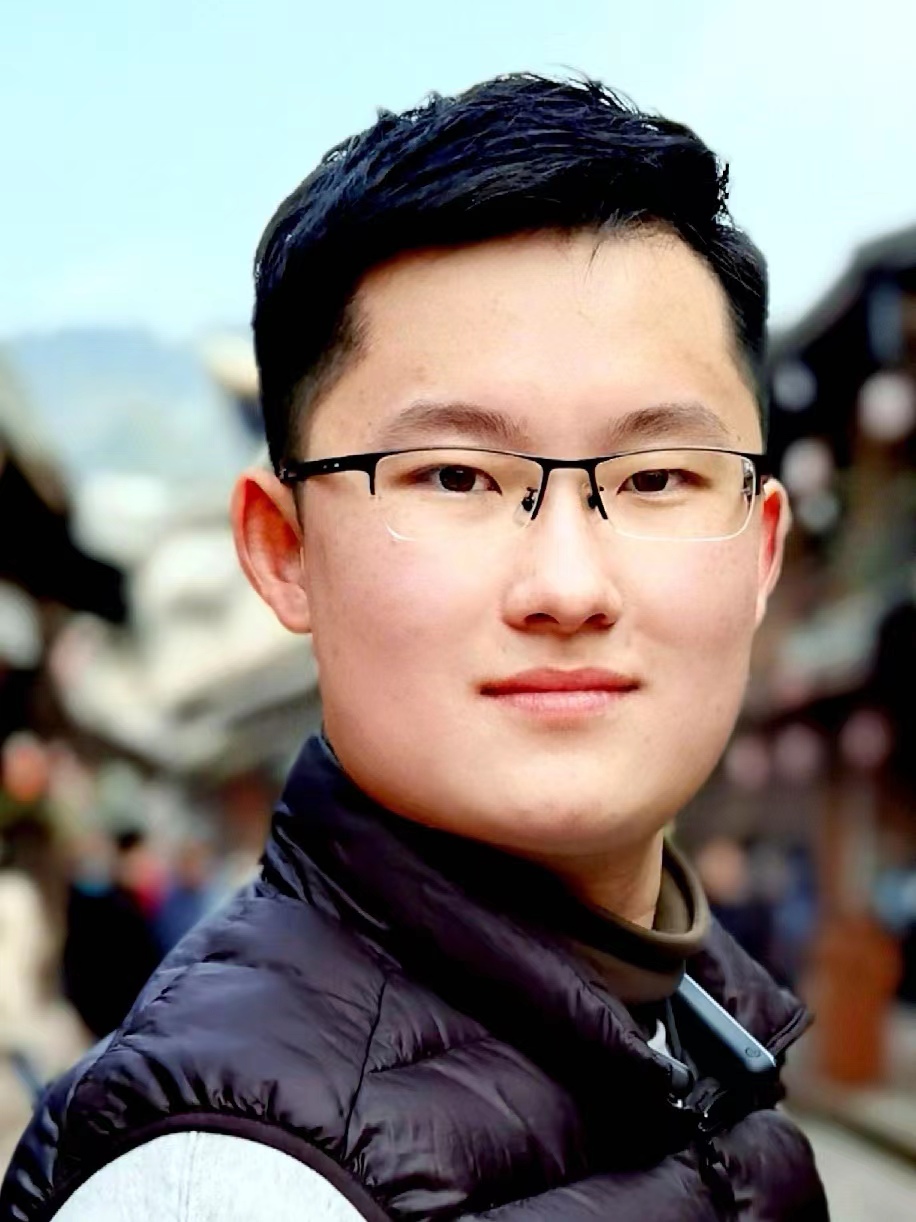}}]{Weicheng Gao}
Weicheng Gao (Graduate Student Member, IEEE), received his B.S. degree in Beijing Institute of Technology in 2022. He is pursuing the Ph.D. degree at the Research Lab of Radar Technology, BIT. He is selected as a member of the China Association for Science and Technology (CAST) Talent Program.\par
His research interests are mainly focused on mathematical theory of radar signal processing and through-the-wall radar human activity recognition.\par
\end{IEEEbiography}
\begin{IEEEbiography}[{\includegraphics[width=1in,height=1.25in,clip,keepaspectratio]{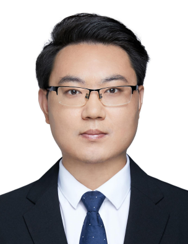}}]{Xiaodong Qu}
Xiaodong Qu (Member, IEEE), associate researcher, received the B.S. degree from Xidian University, Xi’an, China, in 2012, and the Ph.D. degree from the University of Chinese Academy of Sciences, Beijing, China, in 2017.\par
His research interests mainly include array signal processing, through-the-wall radar imaging.\par
\end{IEEEbiography}
\begin{IEEEbiography}[{\includegraphics[width=1in,height=1.25in,clip,keepaspectratio]{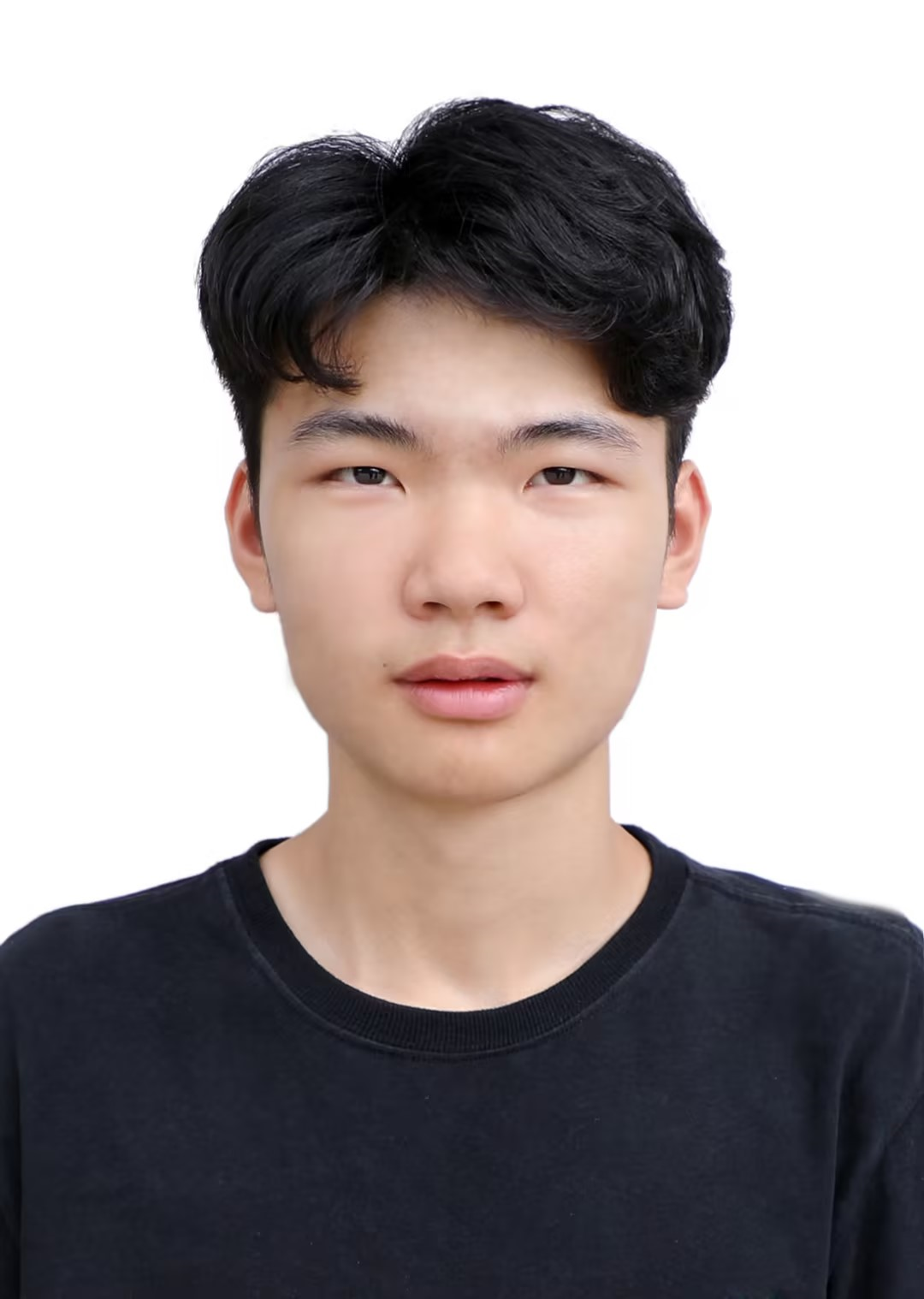}}]{Zeyu Ma}
Zeyu Ma (Student Member, IEEE), received the B.S. degree in automation from North China Electric Power University, Beijing, China, in 2022.  He is currently pursuing the M.S. degree in the School of Information and Electronics at the Beijing Institute of Technology.\par
His research interests include through-the-wall radar moving targets location and tracking.\par
\end{IEEEbiography}
\begin{IEEEbiography}[{\includegraphics[width=1in,height=1.25in,clip,keepaspectratio]{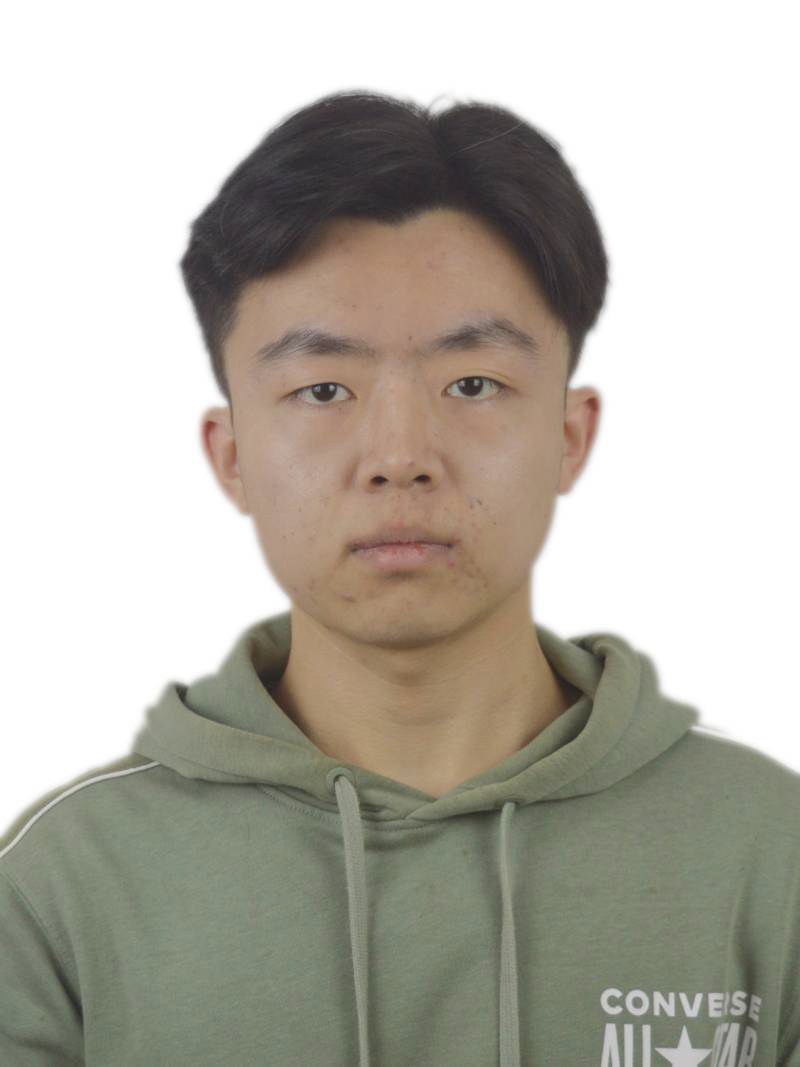}}]{Hao Zhang}
Hao Zhang (Student Member, IEEE), was born in Shanxi Province, China, in 2000. He received the B.S. degree from the North China Electric Power University, Baoding, China, in 2022. He is currently working toward the M.S. degree at the School of Information and Electronics, Beijing Institute of Technology (BIT), Beijing, China.\par 
His current research interests are in the areas of moving target localization with unmanned aerial vehicle based through-the-wall radar.\par
\end{IEEEbiography}

\end{document}